\documentclass[11pt, a4paper]{article}
\pdfoutput=1 

\usepackage{jheppub} 

\usepackage[T1]{fontenc} 

\usepackage[utf8]{inputenc}
\usepackage{epstopdf}
\usepackage[footnotesize]{caption}
\usepackage{amsthm}
\usepackage{enumitem}

\def \be {\begin{equation}}
\def \ee {\end{equation}}
\def \bea {\begin{eqnarray}}
\def \eea {\end{eqnarray}}
\def \nn {\nonumber}

\def \ra {\rangle}
\def \rr {\raise.35ex\hbox{\small $\prime$}\kern-.17em{\mbox{\large $\imath$}}}

\def \dels {\partial\kern-.6em /\kern.1em}
\def \As {{A\kern-.5em / \kern.5em}}
\def \Ds {D\kern-.7em / \kern.5em}

\def \ks {k\kern-.5em /}
\def \ls {l\kern-.5em /}\def \lam {\lambda}

\def \Z {{\mathbb{Z}}}

\def \sgn {\mbox{\small sgn}}

\def \ra {\rightarrow}

\newcommand{\hide}[1]{}
\newcommand*{\sq}{\square}

\newcommand*{\hc}{\dagger}
\newcommand*{\pd}{\partial}

\newcommand*{\rag}{\rangle}
\newcommand*{\lag}{\langle}

\newcommand*{\half}{\frac{1}{2}}

\DeclareMathOperator{\Tr}{Tr}

\newtheorem{theorem}{Theorem}
\newtheorem{definition}{Definition}
\newtheorem{proposition}{Proposition}
\newtheorem{corollary}{Corollary}
\newtheorem{lemma}{Lemma}

\newlist{axioms}{enumerate}{2}
\setlist[axioms,1]{label=\textbf{A\arabic{axiomsi}.}, ref=A\arabic{axiomsi}}
\setlist[axioms,2]{label=\textbf{A\arabic{axiomsi}\rlap{\myEnumCounter{axiomsii}}.},%
                   ref=A\arabic{axiomsi}\myEnumCounter{axiomsii},%
                   align=parleft,%
                   leftmargin=0em,%
                   itemsep=1.4ex,%
                   before={\stepcounter{axiomsi}}}

\title{Entanglement with Centers}

\author[a,1]{Chen-Te Ma \note{Corresponding author.}}

\affiliation[a]{Department of Physics and Center for Theoretical Sciences, 
National Taiwan University, Taipei 10617.}

\emailAdd{yefgst@gmail.com}

\abstract{Entanglement is a physical phenomenon that each state cannot be described individually. Entanglement entropy gives quantitative understanding to the entanglement. We use decomposition of the Hilbert space to discuss properties of the entanglement. Therefore, partial trace operator becomes important to define the reduced density matrix from different centers, which commutes with all elements in the Hilbert space, corresponding to different entanglement choices or different observations on entangling surface. Entanglement entropy is expected to satisfy the strong subadditivity. We discuss decomposition of the Hilbert space for the strong subadditivity and other related inequalities. The entanglement entropy with centers can be computed from the Hamitonian formulations systematically, provided that we know wavefunctional. In the Hamitonian formulation, it is easier to obtain symmetry structure. We consider massless $p$-form theory as an example. The massless $p$-form theory in ($2p+2)$-dimensions has global symmetry, similar to the electric-magnetic duality, connecting centers in ground state. This defines a duality structure in centers. Because it is hard to exactly compute the entanglement entropy from partial trace operator, we propose the Lagrangian formulation from the Hamitonian formulation to compute the entanglement entropy with centers. From the Lagrangian method and saddle point approximation, the codimension two surface term (leading order) in the Einstein gravity theory or holographic entanglement entropy should correspond to non-tensor product decomposition (center is not identity). Finally, we compute the entanglement entropy of the $SU(N)$ Yang-Mills lattice gauge theory in the fundamental representation using the strong coupling expansion in the extended lattice model to obtain spatial area term in total dimensions larger than two for $N>1$.}

\arxivnumber{1511.02671}

\begin{document} 
\maketitle
\flushbottom

\section{Introduction}
\label{1}
Quantum gravity theory with ultraviolet information is expected to be based on the principle of quantum mechanics and gravity theories. The M-theory is a candidate framework of quantum gravity theory by using dualities to unify all fundamental theories. In low-energy limit, we have a suitable low-energy description \cite{Ho:2012dn, Ho:2013paa, Ma:2012hc, Ho:2014una, Ho:2008nn, Zwiebach:1985uq} with duality structures in the M-theory. Ten dimensional low-energy effective theory has the T-duality and S-duality. The T-duality of closed string theory \cite{Zwiebach:1992ie, Saadi:1989tb} exchanges momentum and winding modes, and the T-duality of open string theory exchanges the Dirichlet and Neumann boundary conditions. The T-duality is not a well-defined map as diffeomorphism in the low-energy effective theory so we have non-single valued fields after we perform the T-duality \cite{Hull:2009mi, Hohm:2010jy, Siegel:1993xq, Siegel:1993th, Hull:2009zb, Hohm:2010pp}. A solution is to sacrifice the gauge symmetry to have a global symmetry structure in double space \cite{Ma:2014ala, Ma:2014vqm, Ma:2014kia, Duff:1989tf, Tseytlin:1990va, Tseytlin:1990nb, Siegel:1983es, Ma:2015yma}. This approach is based on geometric construction from the Courant bracket  \cite{Gualtieri:2003dx, Hitchin:2004ut}. The S-duality is a duality between strong and weak coupling constants. Exchanging the coupling constants lead non-perturbative and strongly coupled issues to the ten dimensional low-energy effective theory. One simple and famous example in the S-duality is the electric-magnetic duality \cite{Ho:2013opa, Ho:2015mfa, Deser:1976iy} in four dimensional electromagnetism. The combination of the T-duality and S-duality is the U-duality, and it is studied in the eleven dimensional supergravity with the manifest meaning \cite{Berman:2010is}.

Duality gives the non-trivial equivalence to our theories and large constraints to construction of quantum gravity theory. Well-defined Quantum gravity theory is believed to be a unique description. More restrictions should lead us to probe fundamental properties of quantum gravity. Thermal entropy gives us information to count degrees of freedom of states. This is also consistent with the duality in the low-energy effective theory. For example, the multiple M2-branes theory has expected $N_m^{3/2}$ scaling law in the large $N_m$ limit if we have $N_m$ M2-branes. Thermal entropy should be a suitable quantity to constraint quantum gravity theory.  

Entanglement entropy has more general meaning than the thermal entropy. If we identify the reduced density matrix as $\exp(-\beta H)$, where $H$ is the Hamitonian and $\beta$ is inverse temperature, and replace partial trace operator by trace operator, we should obtain the thermal entropy from the entanglement entropy. The entanglement entropy has similar scaling laws as in the thermal entropy in ten dimensional supergravity theory. The thermal entropy vanishes at zero temperature, but the entanglement entropy is not. The entanglement entropy should have more applications than the thermal entropy. Unfortunately, definition of the entanglement entropy suffers from local gauge symmetry problem. Because we divide space into two parts to define the entanglement entropy, an entangling surface is hard to guarantee gauge symmetry. In order to define the entanglement entropy in a theory with local gauge symmetry, we need to generalize partial trace operator from tensor product decomposition of the Hilbert space to non-tensor product decomposition of the Hilbert space from the Von-Neumann algebra \cite{Casini:2013rba, Casini:2014aia}. In local quantum field theory, the Von-Neumann algebra does not lose generality to describe our quantum theory. Now this definition is based on defining partial trace operator in mathematical sense. What we computed is still unclear in physics. A complete discussion of mathematical inequalities for information meaning is still necessary.

The entanglement entropy reminds us to use more fundamental ways to understand quantum theory. We usually use the Lagrangian or the Hamitonian formulations to understand quantum theories. Constructing the Hamitonian or Lagrangian densities is not an easy work for some gauge symmetries, for example, the multiple M5-branes theory. Many properties in the entanglement entropy can be understood from the Hilbert space before we compute explicitly. In fact, we have complete information even if we only have the Hilbert space. The equivalence between two theories is often checked by duality. A more direct and rigorous way is to use algebras and $n$-point functions in the Hilbert space. Algebraic structure of the entanglement entropy already has useful discussion from the Von-Neumann algebra. How to use algebra to obtain the entanglement entropy should be important in algebraic quantum field theory.

Quantum field theory has one interesting and unsolved strongly coupled problem that we want to understand from various approaches. Our understanding in quantum field theories is almost based on perturbative calculation, only valid in weak coupling region. We do not have generic exact methods to know quantum field theory. In strong coupling regime, confinement is a famous phenomenon in the quantum chromodynamics (QCD). Due to asymptotic freedom behavior in the QCD model from perturbation, we believe that strong coupling regime is in the low-energy domain. Confinement phenomenon leads many theoretical physicists to work in strongly coupled field theory. Even if what we consider is not QCD, we usually expect that similar properties in the QCD can be obtained or a new computation tool will be developed to let us know more for strongly coupled physics. In the M-theory, the $S$-duality or electric-magnetic duality are examples to find strongly coupled physics from weakly coupled regime. This is why many people are interested in duality structure because it sheds light on strongly coupled issues. In the holograph approach, we use anti-de sitter ($AdS$) spacetime to get strongly coupled conformal field theory ($CFT$), which connects gravity theory in weakly coupled regime to the $SU(N)$ super Yang-Mills theory in the adjoint representation in strongly coupled regime from the large $N$, decoupling and near horizon limits in the $AdS_5/CFT_4$ \cite{Maldacena:1997re}. The holograph principle is a conjecture, but there are many evidences to let us be interested in finding some behaviors of physics first from the holograph principle. Standard computation techniques in the entanglement entropy are replica trick and conical method. The computation is very hard to obtain exact solutions in interaction field theory so development of the holograph principle in the entanglement entropy is interesting for understanding behavior of the entanglement entropy before we use field techniques to exactly compute. Because the entanglement entropy does not vanish at zero temperature, it has potential to do order parameter to classify confinement. The entanglement entropy is possibly useful in strongly coupled regime. The study of the holographic entanglement entropy \cite{Casini:2011kv, Maldacena:2013xja, Ryu:2006bv} should help us know more about the entanglement entropy in strongly coupled domain. But this approach is not useful in QCD. The $SU(N)$ super Yang-Mills has many different properties which are not the same as in the QCD model. We do not expect that the holograph principle helps us to understand the QCD model. For a direct computation, it is more convenient to use strong coupling expansion on lattice to compute the QCD model. The strong coupling expansion is a convergent expansion with respect to strong coupling constant. This gives exact results to the strong coupling physics order by order, but the drawback is continuum limit in strong coupling regime. But the confinement can be obtained from the strong coupling expansion in QCD so we believe that some behaviors still hold even if we lose continuum limit. The entanglement entropy in the QCD model can also be computed from the strong coupling expansion on lattice. The difficulties are to overcome gauge symmetry problem on entangling surface. The approach is to define the entanglement entropy from extending the Hilbert space \cite{Donnelly:2011hn}. Other methods of the entanglement entropy on lattice are studied on \cite{Aoki:2015bsa, Soni:2015yga}.

Our goal in this paper is to obtain more mathematical properties and give useful computation methods in the entanglement entropy with center. Now we generalize partial trace operator from the case of tensor product to non-tensor product decomposition of the Hilbert space in the context of the Von-Neumann algebra. Different choices of the centers can be seen as different observations on entangling surface. It is nature to see different entanglement information from different centers. But physical part of the entanglement entropy in quantum field theory should be universal part. The universal part of the entanglement entropy possibly not be affected by a choice of centers on continuous space. The physical interpretation of the entanglement entropy with centers is still unclear. A direct examination is to check mathematical inequalities with information meaning. Even if we do not find these inequalities hold in general situations, it is still interesting to let us understand what decomposition of the Hilbert space gives information meaning. Given a wavefunctional, the Hamitonian formulation is easier to obtain some exact properties. An interesting problem in the center is to find symmetry structure to connect them. This motivation lets us consider a massless $p$-form theory in $(2p+2)$-dimensions. We find global symmetry, similar with the electric-magnetic duality, connecting different choices of the entanglement entropy on a rotation. The exact solution of the entanglement entropy in the Hamitonian formulation relies on numerical studies so we propose the Lagrangian formulation from the Hamitonian formulation. The Lagrangian formulation is easier to compute due to avoiding to use partial trace operator. From the Lagrangian formulation, we find that computation of the entanglement entropy with center is equivalent to considering the Lagrangian with boundary conditions. The Einstein gravity theory has diffeomorphism gauge symmetry and has many similar properties with the non-abelian gauge theory. We expect that the Einstein gravity theory should suffer from the same issue as in gauge theory. We use saddle point approximation to consider the entanglement entropy at leading order. The codimension two surface will be obtained from an entangling surface. This should correspond to non-tensor product decomposition of the Hilbert space. We also check that our computation is also compatible with holograph limit. Finally, we use the extended lattice model \cite{Donnelly:2011hn} to consider the entanglement entropy in the $SU(N)$ lattice gauge theories in the fundamental representation using the strong coupling expansion. We find that the entanglement entropy will vanish in the infinite strong coupling constant so this result possibly confirm color confinement in low-energy domain. 

We discuss various mathematical properties of the entanglement entropy with centers in Sec. \ref{Entanglement Entropy with Center}. We also discuss the massless $p$-form theory in the Hamitonian formulation, propose the Lagrangian formulation to compute the entanglement entropy with center from the Hamitonian formulation, and discuss the decomposition of the Hilbert space in the Einstein gravity theory in Sec. \ref{5}. Then we enlarge the Hilbert space to compute an electric choice of the entanglement entropy in the $SU(N)$ lattice gauge theory in the fundamental representation, and strongly coupled region using strong coupling expansion from an extended lattice model formulation, and discuss results in Sec. \ref{6}. Finally, we conclude in Sec. \ref{7}. We define information in appendix \ref{app1} and review the Von-Neumann algebra in Appendix \ref{app2}. The details of the Strong subadditivity is shown in Appendix \ref{app3}, explicit computation of quantum entropy in free theory in Appendix \ref{app4} and the details of the entanglement entropy in the Einstein gravity is also shown in Appendix \ref{app5}. We also introduce the $AdS_5$ metric in Appendix \ref{app6}, the Hamitonian formulation in the lattice $SU(N)$ Yang-Mills gauge theory in the fundamental representation in Appendix \ref{app7} and review the extended lattice model in Appendix \ref{app8}.


\section{Entanglement Entropy with Center}
\label{Entanglement Entropy with Center}
We use center, which commutes with other elements in the Hilbert space, to define  decomposition of the Hilbert space. In other words, we use center to classify the entanglement information. We expect that the entanglement information should satisfy strong subadditivity and other inequalities or relations so we discuss these mathematical properties.
\subsection{Center}
Quantum properties are hidden in canonical relations. A complete discussion in the entanglement entropy should start from canonical relations. The entanglement entropy is defined from information. The detailed discussion of information is in Appendix \ref{app1}. The classical Shannon entropy is defined as $H_C=-\sum_i p_i\ln p_i$, where we denote space indices from $i$ to $z$, and quantum entropy as $S_Q(\rho)\equiv -\mbox{Tr}\big(\rho\ln\rho\big)$, where $\rho$ is a density matrix and $\mbox{Tr}$ is a trace operator. We also define $0\ln 0\equiv0$ in the entropy quantities.
\begin{definition}
A density matrix $\rho$ on the Hilbert space is a self-adjoint non-negative trace class operator whose trace is unity.
\end{definition}
We have operator algebras on two regions ($V$ and $\bar{V}$), and consider local quantum field theory, then we use algebra 
\bea
A_V=A_{\bar{V}}^{\prime}, \qquad A_{\bar{V}}=A_V^{\prime}
\eea
without losing generality, where $A_V$ is a algebra in region $V$ and $A_{\bar{V}}$ is a algebra in region $\bar{V}$. $A^{\prime}$ is the commutant of $A$.
 If we consider tensor product decomposition of total Hilbert space ($H=H_V\otimes H_{\bar{V}}$), the algebra can be amplified ($A_V\rightarrow A_V\otimes I_{\bar{V}}, A_{\bar{V}}\rightarrow I_V\otimes A_{\bar{V}}$). Then this decomposition corresponds to the trivial choice (center is the identity operator). In general, we should include non-trivial centers (center is not the identity operator) in our discussion. A theory has different centers using different decompositions of the Hilbert space. For example, 
\bea
\lbrack \pi(x), \pi(y)\rbrack=0, \qquad \lbrack \pi(x), \phi(y)\rbrack= -i\delta^{D-1}(x-y), \qquad \lbrack \phi(x), \phi(y)\rbrack=0.
\eea 
This is a standard scalar field theory in $D$-dimensions. In this case, we can remove $\pi$ on entangling surface to let $\phi$ be a center on entangling surface. This choice does not correspond to a tensor product decomposition of the Hilbert space. Removing operators on entangling surface can be seen as a different observation in the local operators on entangling surface. Different observation gives different entanglement information. This ambiguity of the entanglement entropy is generic, and should not give any non-physical issues from this aspect. 

Computing the entanglement entropy with center is to find a basis to diagonalize center. The mathematical proof is in Appendix \ref{app2} \cite{von}, especially for the Theorem \ref{diag1} and the Lemma \ref{diag2}. We first give a procedure of computing for discrete measure, then we discuss results of the continuous measure. A center ($Z$) is isomorphic to
\begin{equation} 
\left(\begin{array}{cccc}
\lambda^11 & 0 & \hdots & 0 \\
0 & \lambda^21 & \hdots & 0 \\
\vdots & \vdots & &\vdots \\
0 & 0 & \hdots & \lambda^m 1
\end{array}\right)\,.
\end{equation}   
Total algebras 
\bea
A\cup A^{\prime}=\bigg(A\cup A^{\prime}\bigg)^{\prime\prime}=\bigg(A^{\prime}\cap A\bigg)^{\prime}=Z^{\prime}
\eea
are isomorphic to
\bea
\left(\begin{array}{cccc}
A_1\otimes A_1^{\prime} & 0 & \hdots & 0 \\
0 & A_2\otimes A_2^{\prime} & \hdots & 0 \\
\vdots & \vdots & &\vdots \\
0 & 0 & \hdots & A_m\otimes A_m^{\prime}
\end{array}\right)\ .
\eea
The algebra $A$ is also isomorphic to a block diagonal form
\bea
\left(\begin{array}{cccc}
A_1 & 0 & \hdots & 0 \\
0 & A_2 & \hdots & 0 \\
\vdots & \vdots & &\vdots \\
0 & 0 & \hdots & A_m
\end{array}\right)\,.
\eea
Therefore, the total Hilbert space ($H$) is isomorphic to $\bigoplus_k \bigg(H_V^k\otimes H_{\bar{V}}^k\bigg)$. We can define a partial trace operator to trace over region $\bar{V}$. Hence, the reduced density matrix in region $V$ is
\bea
\mbox{Tr}_{\bar{V}}\rho_{A_VA_{\bar{V}}}=\rho_{A_V}=
\left(\begin{array}{cccc}
p_1\rho_{A_1} & 0 & \hdots & 0 \\
0 & p_2\rho_{A_2} & \hdots & 0 \\
\vdots & \vdots & &\vdots \\
0 & 0 & \hdots & p_m\rho_{A_m}
\end{array}\right)\,,
\eea
where $\mbox{Tr}\rho_{A_k}=1$ and $\mbox{Tr}_{\bar{V}}$ means that we partial trace over $\bar{V}$. Then, we compute the entanglement entropy $S_{\mbox{EE}}(A)\equiv -\mbox{Tr}\big(\rho_A\ln\rho_A\big)$,
\bea
-\mbox{Tr}\big(\rho_A\ln\rho_A\big)=-\sum_k\mbox{Tr}\big(p_k\rho_{A_k}\ln(p_k\rho_{A_k})\big)
=-\sum_k p_k\ln p_k-\sum_k\mbox{Tr}\big(p_k\rho_{A_k}\ln\rho_{A_k}\big).
\nn\\
\eea
The first term is the classical Shannon entropy and the second term is the average entanglement entropy. We can find that the results are larger than zero even if we have the non-trivial centers. But if we consider continuous distributions, the classical Shannon entropy becomes
\bea
-\sum_{\phi} \big(f(\phi)\Delta\big)\ln(f(\phi)\Delta)\longrightarrow-\ln(\Delta)-\int d\phi\ f(\phi)\ln f(\phi),
\eea
where we replace $p_k$ by $f(\phi)\Delta$.
Then the classical Shannon entropy depends on $\Delta$ or a regulator. Thus, the classical Shannon entropy will depend on regularization scheme. The second term in the classical Shannon entropy does not guarantee positive. In continuous distribution, we possibly find negative term in quantum field computation. The second term in the classical Shannon entropy is called continuous entropy. If we define the entanglement entropy after we remove the regulator or the first term in the classical Shannon entropy, it is called continuous entanglement entropy. To avoid the regulator to appear in our computation, we consider the mutual information \big($M(A, B)\equiv S_{\mbox{EE}}(A)+S_{\mbox{EE}}(B)-S_{\mbox{EE}}(A\cup B)$\big). Because the mutual information should have information meaning, it should increase with degrees of freedom of algebra, and have finite value. Degrees of freedom of algebra should increase as increasing lattice size. If the maximum degree of freedom of algebra is trivial choice,  we expect that the mutual information with the non-trivial centers (removing some operators) should converges to the mutual information with the trivial center. This argument is only valid for a quantity with information meaning. The continuous entanglement entropy possibly not be valid.

\subsection{Properties of the Entanglement with Center}
The entanglement with non-trivial centers is unclear in physical interpretation. One way is to check their theoretical properties. We study partial trace operator and the strong subadditivity or other inequalities with arbitrary centers. If non-trivial choices can capture information, we should find similar results with the trivial choice. In this section, we will show their properties on the discrete space. The extension from discrete space to continuous space is straightforward from replacing discrete distribution by continuous distribution.

\subsubsection{Partial Trace Operator}
The partial trace operator \cite{Araki:1970ba} is important to define the reduce density matrix from the density matrix to extract the entanglement entropy. When we consider the entanglement entropy with non-trivial centers, the partial trace operator should be generalized from the case of the trivial choice. 

\begin{definition}
The density matrix $\rho$ is a pure state if $\rho$ is a projection operator onto an one-dimensional subspace. In other words, $\rho x=y\lag y, x\rag$ with $|y|=1$, where $\lag y, x\rag$ is the inner product space between $x$ and $y$.
\end{definition}

\begin{lemma}
\label{partial trace operator}
Let $\rho_{12}$ be a pure state density matrix on $H_{12}$ isomorphic to $\bigoplus_k H_1^k\otimes H_2^k$. Let $f( \cdot )$ be a real valued function, and $f(0)=0$. Then we obtain
\bea
\mathrm{Tr}\ f(\rho_1)=\mathrm{Tr}\ f(\rho_2).
\eea
 In particular, $S_1=S_2$.
\end{lemma}
\begin{proof}[Proof]
Let $\rho_{12}x=\lag y,x\rag y$, $y=\sum_{k,i}\lambda_i^k y_{1i}^k\otimes y_{2i}^k$, where $y_{1i}^k$ and $y_{2i}^k$ are orthonormal. Let $P(y_{\alpha i}^k)x \equiv\lag y_{\alpha i}^k, x\rag y_{\alpha i}^k$ be the projection on the one-dimensional subspace of $H_{\alpha}^k$ which contains $y_{\alpha i}^k$. Then $\rho_{\alpha}x=\sum_{i,k}(\lambda_i^{k})^2P(y_{\alpha i}^k)x$. Hence, $\rho_1$ and $\rho_2$ have the same eigenvalues and multiplicities except for zero eigenvalues. Therefore, we obtain
\bea
\label{pto}
\mbox{Tr}\ f(\rho_1)=\mbox{Tr}\ f(\rho_2). 
\eea
It is direct to deduce $S_1=S_2$.
\end{proof}
This lemma shows that the entanglement entropy with a generic center does not change $S_1=S_2$ when the density matrix is a pure state. The partial trace operator is an ambiguous operator. All ambiguities of the entanglement entropy come from how to define this operator. However, a strong evidence in \eqref{pto} for the partial trace operator still has good properties from its generalization even if we consider non-trivial centers because the form of the relation \eqref{pto} does not modify from arbitrary real valued functions. 

\begin{lemma}
\label{epto}
Let $\rho_{1}$ be a reduced density matrix on $\bigoplus_k H_1^k$. Then there exists a pure state density matrix $\rho_{12}$ on $H_{12}$ isomorphic to $\bigoplus_k H_1^k\otimes H_2^k$ such that
\bea
\mathrm{Tr}_2\ \rho_{12}=\rho_1.
\eea
\end{lemma}
The proof of the Lemma \ref{epto} is similar with the Lemma \ref{partial trace operator}, and it establishes that a reduced density matrix can exist correspondent density matrix and partial trace operator. This mathematical property enhances that the entanglement entropy with non-trivial centers possibly have the similar properties with the entanglement entropy of the trivial choice. However, we will give more properties to the entanglement entropy with the non-trivial centers to understand physical implications. 

\subsubsection{Decomposition of the Hilbert Space}
The decomposition of the Hilbert space is a subtle issue in the entanglement entropy. A suitable decomposition is to offer a proper partial trace operator in a reduced density matrix to obtain the entanglement entropy. The fist case is 
\bea
H_1\otimes H_2\otimes\cdots
\eea
and the center is
\bea
Z_1\otimes Z_2\otimes\cdots.
\eea
This example is more general than the trivial choice. The Hilbert space does not have any problems to choose tensor product decomposition if you do not guarantee local symmetry on the entangling surface. When we consider the Von-Neumann algebra to do decomposition, the reason comes from the non-trivial centers on the entangling surface. Here, we do not necessary need the properties of the Von-Neumann algebra to help us to decompose our Hilbert space.
The second example is to consider algebra 
\bea
A_1^{\prime}=A_2\cup A_3, \qquad A_2^{\prime}=A_1\cup A_3, \qquad A_3^{\prime}=A_1\cup A_2
\eea
 and the center is
\bea
A_1\cap A_2=1, \qquad A_1\cap A_3=Z_1, \qquad A_2\cap A_3=Z_2.
\eea
Then total Hilbert space has a tensor product structure as
\bea
H_{123}&\sim&\bigoplus_k H_1^k\otimes H_{2}^k\otimes H_{3}.
\eea
For more difficult examples, we possibly not have this kind of complete tensor product structure. But we can define entanglement entropy. For example, we can find center to decompose one total Hilbert space to two separate Hilbert spaces when we consider local quantum field theory. This is enough for us to define the entanglement entropy, but the entanglement entropy may not be defined in the same basis. 

But you may be confused why we cannot define the entanglement entropy in the same basis if our centers commute with each other. Let us use the second example to interpret more on this point. We can obtain an isomorphic Hilbert space $H_{12}\otimes H_3$ using $Z_1$ and $Z_2$. Thus, we find a way to define the entanglement entropy in the same basis. But it is only useful when you consider entanglement in two regions. If you want to partial trace over regions one or two, then the center between regions one and two needs to be identity if you consider the Von-Neumann algebra. Of course, you can argue that we can remove some operators between regions one and two to define the reduced density matrix or the entanglement entropy. But you do not use the same total Hilbert space to define the entanglement.  You may also argue that we can remove operators between regions one and three, and regions two and three in the total Hilbert space first. Then you do not know how to perform partial trace to get $\rho_{12}$ because the algebra is not the Von-Neumann algebra.

 Hence, the entanglement entropy is not defined in the same Hilbert space. If your Hilbert space is changed, then you also change your basis to detect the entanglement entropy. The Von-Neumann algebra has a simple structure to decompose the Hilbert space with non-trivia centers, but it only gives us entanglement in two regions. 

Our mathematical proof in the strong subadditivity possibly suffers from this problem so non-trivial center in the entanglement entropy do not have clear physical interpretation now.  We also remind that removing operators should change total Hilbert space and wavefunctional. But if you do not observe operators that you removed, they will give the same observable. This is a way to extract the entanglement information from the partial trace operator, but we need to let some states be classical states. Although we lose complete information, we obtain entanglement information.

\subsubsection{Entanglement Inequalities}
We start to discuss the strong subadditivity \cite{Araki:1970ba, Lieb:1973cp} and other related inequalities. These inequalities will give more information interpretation and point out more problems to the entanglement entropy with center. We give all necessary details \cite{Araki:1970ba, Lieb:1973cp} in Appendix \ref{app3}.

\begin{theorem}
\label{strong subadditivity}
\bea
S_{123}+S_2-S_{12}-S_{23}\le 0.
\eea
\end{theorem}
\begin{proof}[Proof]
We use the Lemma \ref{Klein inequality} with $A=\rho_{123}$ and $B=\exp(-\ln\rho_2+\ln\rho_{12}+\ln\rho_{23})$ to find
\bea
F(\rho_{123})=S_{123}+S_2-S_{12}-S_{23}\le\mbox{Tr}\bigg(\exp(\ln\rho_{12}-\ln\rho_2+\ln\rho_{23})-\rho_{123}\bigg),
\eea
and apply the Theorem \ref{Golden-Thompson inequality} to obtain
\bea
\mbox{Tr}\bigg(\exp(\ln C-\ln D+\ln E)\bigg)\le\mbox{Tr}\bigg(\int_0^{\infty}dx\ C(D+x1)^{-1}E(D+x1)^{-1}\bigg).
\eea
Hence, we obtain 
\bea
F(\rho_{123})&\le&\mbox{Tr}\bigg(-\rho_{123}+\int_0^{\infty}dx\ \rho_{12}(\rho_2+x1)^{-1}\rho_{23}(\rho_2+x1)^{-1}\bigg)
\nn\\
&=&-\mbox{Tr}\ \rho_{123}+\mbox{Tr} \bigg(\int_0^{\infty}dx\ \rho_2(\rho_2+x1)^{-1}\rho_2(\rho_2+x1)^{-1}\bigg)
\nn\\
&=&\mbox{Tr}\ \rho_2-\mbox{Tr}\ \rho_{123}=0
\eea
when we use $C=\rho_{12}$, $D=\rho_2$ and $E=\rho_{23}$.
We used 
\bea
\mbox{Tr}_{13}\ \rho_{23}=\rho_2, \qquad \mbox{Tr}_{13}\ \rho_{12}=\rho_2
\eea
in the first equality. This is a very subtle place. If we do not use the same basis to define the entanglement entropy, we do not have the first equality. This shows that the strong subadditivity is not valid for generic centers on entangling surface. 
\end{proof}
If you can use the same Hilbert space or basis to define the entanglement entropy, the strong subadditivity still holds. Unfortunately, gauge theory needs the non-trivial centers to define the entanglement with gauge symmetry. The violation of the strong subadditivity \cite{Casini:2014aia} must come from this reason because our mathematical proof only uses
\bea
\mbox{Tr}_{13}\ \rho_{23}=\rho_2, \qquad \mbox{Tr}_{13}\ \rho_{12}=\rho_2, \qquad \mbox{Tr}_{3}\ \rho_{123}=\rho_{12}, \qquad \mbox{Tr}_{1}\ \rho_{123}=\rho_{23}.
\eea

Now we discuss positivity of the mutual information via the strong subadditivity.
\begin{corollary}
If $\rho_{12}$ is a density matrix on $H_{12}$ which is isomorphic to $\bigoplus_k H_1^k\otimes H_2^k$, then
\bea
S_{12}\le S_1+S_2.
\eea
\end{corollary}
\begin{proof}[Proof]
From the Theorem \ref{strong subadditivity}, we have 
\bea
\lag S_{123}\rag+\lag S_2\rag\le \lag S_{12}\rag+\lag S_{23}\rag
\eea
on $\bigg(\bigoplus_k H_1^k\otimes H_2^k\bigg)\otimes H_3$.
Interchanging $2$ and $3$ and take $H_3$ be one dimension. Then we obtain
\bea
\lag S_{12}\rag\le \lag S_1\rag+\lag S_2\rag.
\eea
Because the classical Shannon entropy also has this relation (Its proof is similar with the quantum entropy.), we get
\bea
S_{12}\le S_1+S_2.
\eea
\end{proof}
The mutual information
\bea
M(A, B)=\lag S_A\rag+\lag S_B\rag-\lag S_{AB}\rag+
\sum_{k_A, k_B}p_{k_A, k_B}\ln\bigg(\frac{p_{k_A, k_B}}{p_{k_A}p_{k_B}}\bigg)
\eea
 is still positive when we consider the total Hilbert space isomorphic to $\bigoplus_k H_1^k\otimes H_2^k$. The strong subadditivity is important for us to guarantee positivity for the mutual information. If we lose the the strong subadditivity, we need to be careful about physical interpretation. 

Finally, we try to rewrite the strong subadditivity as before. From the Lemma \ref{partial trace operator} ($S_{123}=S_4$, $S_{12}=S_{34}$), the strong subadditivity becomes
\bea
S_4+S_2\le S_{34}+S_{23}.
\eea
Then we replace $4$ by $1$. Hence, we can rewrite the strong subadditivity
\bea
S_1+S_2\le S_{13}+S_{23}.
\eea
We remind that this strong subadditivity is correct in the trivial choice. Because we use $S_{123}=S_4$, $S_{12}=S_{34}$, we need to let center be identity in regions one, two and three. This also gives us a restriction to the center in the region four because we replace index $4$ by $1$.

\section{Computation Methods in the Entanglement Entropy with Center}
\label{5}
We will show the Hamitonian formulation and discuss global symmetry structure of a massless $p$-form theory in $(2p+2)$-dimensions. Based on the Hamitonian formulation, we construct the Lagrangian method for computing the entanglement entropy with center. Finally, we use this Lagrangian formulation to discuss decomposition of the Hilbert space in the Einstein gravity theory.

\subsection{The Hamitonian Formulation in a $p$-Form Theory}
We show the Hamitonian method \cite{Casini:2013rba, Casini:2014aia, Casini:2009sr} for a $p$-form theory. We first discuss scalar field theory (0-form theory). Then we extend our discussion to the abelian $p$-form Yang-Mills gauge theory, and consider the canonical momentum $p_i$ and position operators $q_i$, $i\in V=\{1, 2,\cdots, n\}$. When we compute the entanglement entropy with centers, we can choose the subset of the momentum $p_{i_{B}}$, $i_B\in B=\{k+1, k+2,\cdots, n\}$. Then the center is $q_{i_A}$ $i_A\in A=\{1, 2, \cdots, k\}$. We indicate the indices in $A$ from $i_A$ to $z_A$ and the indices in $B$ from $i_B$ to $z_B$. Finally, we find that a massless $p$-form theory has  an equivalent choices of the entanglement entropy from a global rotation symmetry in $(2p+2)$-dimensions.

\subsubsection{Scalar Field Theory}
We consider scalar field theory with the Hamitonian 
\bea
H_{SF}(q, p)=\frac{1}{2}\bigg(\sum_i p_i^2+\sum_{i, j}q_iM_{i, j}q_j\bigg),
\eea
where $q\equiv q_i$, $p\equiv p_i$ and $M\equiv M_{ij}$ is not related to field, and is symmetric. The commutation relations are
\bea
\lbrack q_i, p_j \rbrack=i\delta_{ij}, \qquad \lbrack q_i, q_j\rbrack=0, \qquad 
\lbrack p_i, p_j\rbrack=0
\eea
and the density matrix is
\bea
\rho(q, q^{\prime})=C_1\exp\bigg\lbrack-\frac{1}{2}\bigg(qM^{\frac{1}{2}}q+q^{\prime}M^{\frac{1}{2}}q^{\prime}\bigg)\bigg\rbrack,
\eea
where $C_1$ is a normalization constant and $q^{\prime}\equiv q^{\prime}_i$. The expectation value is defined as
\bea
\lag O(q, p)\rag=\int dq\ \bigg|O(q, -i\partial_q)\rho(q, q^{\prime})\bigg|_{q=q^{\prime}}.
\eea
Hence, two point functions are
\bea
\lag q_iq_j\rag\equiv X_{ij}=\frac{1}{2}M^{-\frac{1}{2}}_{ij}, \qquad\lag p_ip_j\rag\equiv P_{ij}=\frac{1}{2}M^{\frac{1}{2}}_{ij},\qquad \lag q_ip_j\rag=\frac{i}{2}\delta_{ij}
\eea
Now we let $i={i_A}\oplus {i_B}$ and $q^A=q^{\prime A}=\tilde{q}^A$, where $q^A\equiv q_{i_A}$, $q^{\prime A}\equiv q^{\prime}_{i_A}$ and $\tilde{q}^A\equiv \tilde{q}_{i_A}$. Because we want to partial trace over region $A$, and have centers in the total Hilbert space, we choose a particular value $\tilde{q}^A$ for $q^A$ and $q^{\prime A}$. Therefore, the density matrix is given by
\bea
\rho(q, q^{\prime})&=&C_2\exp\bigg\lbrack-\frac{1}{2}\bigg(2\tilde{q}^AM^{\frac{1}{2}}_{AA}\tilde{q}^A
+q^BM^{\frac{1}{2}}_{BB}q^B+2\tilde{q}^AM^{\frac{1}{2}}_{AB}q^B
+q^{\prime B}M^{\frac{1}{2}}_{BB}q^{\prime B}+2\tilde{q}^AM^{\frac{1}{2}}_{AB}q^{\prime B}\bigg)\bigg\rbrack
\nn\\
&=&C_2\exp\bigg\lbrack-\frac{1}{2}\begin{pmatrix} \tilde{q}^A &q^B \end{pmatrix}
\begin{pmatrix} M_{AA}^{\frac{1}{2}} & M^{\frac{1}{2}}_{AB}
 \\ M_{BA}^{\frac{1}{2}} & M_{BB}^{\frac{1}{2}} \end{pmatrix}
\begin{pmatrix} \tilde{q}^A
 \\ q^B \end{pmatrix}
-\frac{1}{2}\begin{pmatrix} \tilde{q}^A &q^{\prime B} \end{pmatrix}
\begin{pmatrix} M_{AA}^{\frac{1}{2}} & M^{\frac{1}{2}}_{AB}
 \\ M_{BA}^{\frac{1}{2}} & M_{BB}^{\frac{1}{2}} \end{pmatrix}
\begin{pmatrix} \tilde{q}^A
 \\ q^{\prime B} \end{pmatrix}
\bigg\rbrack,
\nn\\
\eea
where $C_2$ is a normalization constant, $q^B\equiv q_{i_B}$, $q^{\prime B}\equiv q^{\prime}_{i_B}$, $M_{AA}\equiv M_{i_Ai_A}$, $M_{BB}\equiv M_{i_Bi_B}$ and $M_{AB}\equiv M_{i_Ai_B}$. From
\bea
&&
\begin{pmatrix} \tilde{q}^A &q^B \end{pmatrix}
\begin{pmatrix} M_{AA}^{\frac{1}{2}} & M^{\frac{1}{2}}_{AB}
 \\ M_{BA}^{\frac{1}{2}} & M_{BB}^{\frac{1}{2}} \end{pmatrix}
\begin{pmatrix} \tilde{q}^A
 \\ q^B \end{pmatrix}
\nn\\
&=&\begin{pmatrix} \tilde{q}^A &q^B \end{pmatrix}
\begin{pmatrix} 1 & M^{\frac{1}{2}}_{AB}(M_{BB})^{-\frac{1}{2}}
 \\ 0 & 1 \end{pmatrix}
\begin{pmatrix} 1 & -M^{\frac{1}{2}}_{AB}(M_{BB})^{-\frac{1}{2}}
 \\ 0 & 1 \end{pmatrix}
\begin{pmatrix} M_{AA}^{\frac{1}{2}} & M^{\frac{1}{2}}_{AB}
 \\ M_{BA}^{\frac{1}{2}} & M_{BB}^{\frac{1}{2}} \end{pmatrix}
\nn\\
&&\cdot
\begin{pmatrix} 1 & 0
 \\ -(M_{BB})^{-\frac{1}{2}}M_{BA}^{\frac{1}{2}} & 1 \end{pmatrix}
\begin{pmatrix} 1 & 0
 \\ (M_{BB})^{-\frac{1}{2}}M_{BA}^{\frac{1}{2}} & 1 \end{pmatrix}
\begin{pmatrix} \tilde{q}^A
 \\ q^B \end{pmatrix}
\nn\\
&=&
\begin{pmatrix} \tilde{q}^A &q^B+\tilde{q}^AM_{AB}^{\frac{1}{2}}(M_{BB})^{-\frac{1}{2}} \end{pmatrix}
\begin{pmatrix}M_{AA}^{\frac{1}{2}}-M_{AB}^{\frac{1}{2}}(M_{BB})^{-\frac{1}{2}}M_{BA}^{\frac{1}{2}} & 0
 \\ 0 & M_{BB}^{\frac{1}{2}} \end{pmatrix}
\begin{pmatrix} \tilde{q}^A
 \\ q^B+(M_{BB})^{-\frac{1}{2}}M_{BA}^{\frac{1}{2}}\tilde{q}^A \end{pmatrix}
\nn\\
\eea
and
\bea
&&
\begin{pmatrix} \tilde{q}^A &q^{\prime B} \end{pmatrix}
\begin{pmatrix} M_{AA}^{\frac{1}{2}} & M^{\frac{1}{2}}_{AB}
 \\ M_{BA}^{\frac{1}{2}} & M_{BB}^{\frac{1}{2}} \end{pmatrix}
\begin{pmatrix} \tilde{q}^A
 \\ q^{\prime B} \end{pmatrix}
\nn\\
&=&\begin{pmatrix} \tilde{q}^A &q^{\prime B} \end{pmatrix}
\begin{pmatrix} 1 & M^{\frac{1}{2}}_{AB}(M_{BB})^{-\frac{1}{2}}
 \\ 0 & 1 \end{pmatrix}
\begin{pmatrix} 1 & -M^{\frac{1}{2}}_{AB}(M_{BB})^{-\frac{1}{2}}
 \\ 0 & 1 \end{pmatrix}
\begin{pmatrix} M_{AA}^{\frac{1}{2}} & M^{\frac{1}{2}}_{AB}
 \\ M_{BA}^{\frac{1}{2}} & M_{BB}^{\frac{1}{2}} \end{pmatrix}
\nn\\
&&\cdot
\begin{pmatrix} 1 & 0
 \\ -(M_{BB})^{-\frac{1}{2}}M_{BA}^{\frac{1}{2}} & 1 \end{pmatrix}
\begin{pmatrix} 1 & 0
 \\ (M_{BB})^{-\frac{1}{2}}M_{BA}^{\frac{1}{2}} & 1 \end{pmatrix}
\begin{pmatrix} \tilde{q}^A
 \\ q^{\prime B} \end{pmatrix}
\nn\\
&=&
\begin{pmatrix} \tilde{q}^A &q^{\prime B}+\tilde{q}^AM_{AB}^{\frac{1}{2}}(M_{BB})^{-\frac{1}{2}} \end{pmatrix}
\begin{pmatrix}M_{AA}^{\frac{1}{2}}-M_{AB}^{\frac{1}{2}}(M_{BB})^{-\frac{1}{2}}M_{BA}^{\frac{1}{2}} & 0
 \\ 0 & M_{BB}^{\frac{1}{2}} \end{pmatrix}
\begin{pmatrix} \tilde{q}^A
 \\ q^{\prime B}+(M_{BB})^{-\frac{1}{2}}M_{BA}^{\frac{1}{2}}\tilde{q}^A \end{pmatrix},
\nn\\
\eea
we can change variables
\bea
q^B\rightarrow q^B-(M_{BB})^{-\frac{1}{2}}(M_{BA})^{\frac{1}{2}}\tilde{q}^A, \qquad q^{\prime B}\rightarrow q^B-(M_{BB})^{-\frac{1}{2}}(M_{BA})^{\frac{1}{2}}\tilde{q}^A
\eea
to rewrite the density matrix, and define probability of the center as
\bea
P(\tilde{q}^A)=\det\bigg(\pi Y_A\bigg)^{-\frac{1}{2}}e^{-\tilde{q}^A(Y_A)^{-1}\tilde{q}^A},
\eea
where $Y_A\equiv Y_{i_A j_A}=\bigg(M_{AA}^{\frac{1}{2}}-M_{AB}^{\frac{1}{2}}(M_{BB})^{-\frac{1}{2}}M_{BA}^{\frac{1}{2}}\bigg)^{-1}=\big((2P)^{-1}\big)_A\equiv\big((2P)^{-1}\big)_{i_Aj_A}$. We used the analytic inversion formula
\bea
\begin{pmatrix} a &b
 \\ c & d \end{pmatrix}^{-1}&=&\begin{pmatrix} a^{-1}+a^{-1}b(d-ca^{-1}b)^{-1}ca^{-1} &-a^{-1}b(d-ca^{-1}b)^{-1}
 \\ -(d-ca^{-1}b)^{-1}ca^{-1} & (d-ca^{-1}b)^{-1} \end{pmatrix}
 \nn\\
 &=&\begin{pmatrix} (a-bd^{-1}c)^{-1} &-(a-bd^{-1}c)^{-1}bd^{-1}
 \\ -d^{-1}c(a-bd^{-1}c)^{-1} & d^{-1}+d^{-1}c(a-bd^{-1}c)^{-1}bd^{-1} \end{pmatrix}
\eea
to relate $Y_A$ to $\big((2P)^{-1}\big)_A$.
Then the density matrix becomes
\bea
\rho(q, q^{\prime})=C_3P(\tilde{q}^A)\exp\bigg\lbrack-\frac{1}{2}\bigg(q^BM_{BB}^{\frac{1}{2}}q^B+q^{\prime B}M_{BB}^{\frac{1}{2}}q^{\prime B}\bigg)\bigg\rbrack.
\eea
The continuous entanglement entropy comes from the combination of the quantum entropy and continuous entropy as 
\bea
S_{CEE}(V)=S_Q(V)+H_{CC}(A).
\eea
The continuous entropy is
\bea
H_{CC}(A)=-\int dq\ P(q^A)\ln P(q^A)=\frac{1}{2}\mbox{Tr}\bigg(1+\ln(\pi Y^A)\bigg).
\eea
If centers live in $A\subseteq V$ and $A^{+} \subseteq V^{+}$, the mutual information between two regions $V$ and $V^{+}$ is given by
\bea
M( V, V^{+})&=&S(V)+S(V^{+})-S(VV^{+})
\nn\\
&=&\frac{1}{2}\mbox{Tr}\bigg(\ln Y_A+\ln Y_{A^+}-\ln Y_{AA^+}\bigg)+S_Q(V)+S_Q(V^+)-S_Q(VV^+).
\nn\\
\eea
The quantum entropy is given by
\bea
S_Q(V)
&=&\mbox{Tr}\bigg((C+\frac{1}{2})\ln(C+\frac{1}{2})-(C-\frac{1}{2})\ln(C-\frac{1}{2})\bigg),
\nn\\
\eea
where $C\equiv(X^BP^B)^{1/2}$, $X^B\equiv X_{i_Bj_B}$ and $P^B\equiv P_{i_Bj_B}$. The explicit computation of quantum entropy is in Appendix \ref{app4}. This result is interesting because it only depends on two point functions. Although this interesting property should only exist in free theories, it possibly gives us some hints to obtain the entanglement entropy from algebraic approach.

\subsubsection{$p$-form Abelian Yang-Mills Gauge Theory}
The computation of the entanglement entropy in the $p$-form abelian Yang-Mills gauge theory is different from the scalar field theory because we need to use the non-canonical variables. The Hamitonian of the $p$-form abelian Yang-Mills gauge theory is
\bea
H_{PAYM}=\frac{1}{2p!}\bigg(F_{0i_1, i_2, \cdots, i_{p}}F_{0i_1, i_2, \cdots, i_{p}}+\frac{1}{p+1}F_{i_1, i_2, \cdots, i_{p+1}}F_{i_1, i_2, \cdots, i_{p+1}}\bigg).
\eea
We can use the multiple indices to write a convenient form as
\bea
H_{PAYM}(q, p)=\frac{1}{2}\bigg(p_{I_S}^2+q_{I^{\prime}_S}q_{I^{\prime}_S}\bigg),
\eea
where $I_S\equiv (i_1, i_2, \cdots, i_{p})$ and $I^{\prime}_S\equiv (i^{\prime}_1, i^{\prime}_2, \cdots, i^{\prime}_{p+1})$ with $i_1>i_2>\cdots >i_{p}$ and $i^{\prime}_1>i^{\prime}_2>\cdots >i^{\prime}_{p+1}$. The commutation relations are given by
\bea
\lbrack q, p\rbrack=iC, \qquad \lbrack q, q\rbrack=0, \qquad \lbrack p, p\rbrack=0.
\eea
We will discuss $C$ later. We introduce the canonical variables ($q^c$ and $p$) from
\bea
q^c=C^{-1}q.
\eea
Because we do not always have equal degrees of freedom between momentum and coordinate operators, we only have right inverse for $C$ generically as
\bea
CC^{-1}=1.
\eea
The Hamitonian becomes
\bea
H_{PAYM}(q^c, p)=\frac{1}{2}\bigg(p^2+q^cM_cq^c\bigg),
\eea
where $M_c=C^TC$. Two point correlation functions are given by
\bea
\lag pp\rag=\frac{1}{2}(C^TC)^{\frac{1}{2}}\equiv P^c, \qquad \lag q^cq^c\rag=\frac{1}{2}(C^TC)^{-\frac{1}{2}}\equiv X^c, \qquad \lag q^cp\rag=\frac{i}{2}1.
\eea
This also gives
\bea
\lag pp\rag=\frac{1}{2}(C^TC)^{\frac{1}{2}}=P^c, \qquad \lag qq\rag=\frac{1}{2}(CC^T)^{\frac{1}{2}}\equiv X, \qquad \lag qp\rag=\frac{i}{2}C.
\eea
The quantum entropy is
\bea
S_Q(V)=\mbox{Tr}\bigg((\theta^c+\frac{1}{2})\ln(\theta^c+\frac{1}{2})-(\theta^c-\frac{1}{2})\ln(\theta^c-\frac{1}{2})\bigg),
\eea
where $\theta^c=(X^c_BP^c_B)^{1/2}$, $X^c_B\equiv X^c_{I_{SB}J_{SB}}$ and  $P^c_B\equiv P^c_{I_{SB}J_{SB}}$. The computation of the quantum entropy is the same as in the free scalar field theory so we get the similar form of quantum entropy with the free scalar field theory.
The continuous entropy is given by
\bea
H_{CC}(A)=\frac{1}{2}\mbox{Tr}\bigg(1+\ln(\pi Y_A)\bigg),
\eea
where $Y_A\equiv Y_{I_{SA}J_{AS}}=\big((2P^c)^{-1}\big)_A$.
The commutation relation related to $C$ of $p$-form abelian Yang-Mills gauge theory in $D$-dimensions ($D>p+1$) is given by 
\bea
\lbrack A_{J_S}(x), F_{0K_S}(x^{\prime})\rbrack=\frac{i}{p!}\eta_{J_SK_S}\delta^{D-1}(x-x^{\prime}),
\eea
where
\bea
\eta_{I_SJ_S}=\sum_{\pi}\sgn(\pi)\eta_{i_{\pi_1}j_1}\eta_{i_{\pi_2}j_2}\cdots\eta_{i_{\pi_p}j_p},
\eea
where $\pi$ is a permutation operation.
In $(2p+2)$-dimensions, we have 
\bea
\lbrack B_{I_S}(x), F_{0J_S}(x^{\prime})\rbrack=i\epsilon_{I_SkJ_S}\partial^k\delta^{2p+1}(x-x^{\prime}),
\eea
where $B_{I_S}=\frac{1}{(p+1)!}\epsilon_{I_S J_Sk}F^{J_S k}$ is magnetic field. The Hamitonian of the abelian Yang-Mills gauge theory of $q$ can be replaced by magnetic field in $(2p+2)$ dimensions. After we use the magnetic field to represent the Hamitonian, we get a correspond $C_{I_SJ_S}=\epsilon_{I_SkJ_S}\partial^k\delta^{2p+1}(x-x^{\prime})$. In the case of 1-form, $C$ is determined by
\bea
\lbrack F_{ii^{\prime}}(x), F_{0j}(x^{\prime})\rbrack=i\bigg(\delta_{ij}\partial_{i^{\prime}}\delta^{D-1}(x-x^{\prime})-\delta_{i^{\prime}j}\partial_{i}\delta^{D-1}(x-x^{\prime})\bigg)=iC_{ii^{\prime}j}.
\eea
We also find one interesting result in $(2p+2)$-dimensions. The electric choice (removing magnetic field) is equivalent to the magnetic choice (removing electric field). Our results are the continuous entanglement entropy with the center $q$. When considering the center $p$, we just exchange $X$ and $P^c$ in the continuous entropy. We find $X=P^c$ in $(2p+2)$-dimensions. This result is expected because we have electric-magnetic duality in $(2p+2)$-dimensions. Therefore, we can conclude the equivalent entanglement entropy can be chosen via 
\bea
\tilde{E}=E\cos\theta+B\sin\theta, \qquad \tilde{B}=E\cos\theta-B\sin\theta,
\eea
where $E$ and $B$ are electric and magnetic fields. 
We use the Hamitonian
\bea
H_{SF2}(q, p)=\frac{1}{2}\bigg(p^2+\partial_1 q \partial_1 q\bigg)
\eea
for massless scalar field theory in two dimensions. The commutation relations are given by
\bea
\lbrack p(x), \partial_1 q(y)\rbrack=-i\partial_1\delta(x-y), \qquad \lbrack \partial_1q(x),\partial_1 q(y)\rbrack=0, \qquad \lbrack p(x), p(y)\rbrack=0.
\eea
From the commutation relations, we also find equivalent entanglement entropy via 
\bea
\tilde{p}\rightarrow p\cos\theta+\partial_1 q \sin\theta, \qquad \partial_1\tilde{q}\rightarrow p\cos\theta-\partial_1q\sin\theta.
\eea
The global symmetry structure in center possibly helps us to find the entanglement entropy with local symmetry to classify centers even if we consider non-trivial centers. In the $\Z_N$ lattice gauge theory, \cite{Radicevic:2014kqa} finds a different duality to relate some choices in different dimensions.  

\subsection{The Lagrangian Formulation}
We propose the Lagrangian method to consider the entanglement entropy with center based on the Hamitonian formulation, and give standard computation methods in the Lagrangian formulation.

\subsubsection{Lagrangian}
The computation of the entanglement entropy in the Lagrangian method is not hard to derive from the Hamitonian formulation. The Hamitonian method is a direct way to compute the entanglement entropy if you have a ground state wavefunctional, and it is easier to understand ambiguities of the entanglement entropy. But the computation related to partial trace operator is hard to get an exact solution of the entanglement entropy. The Lagrangian formulation avoids defining the partial trace operator to compute the entanglement entropy. We start from the path integral representation of ground state wavefunctional
\bea
\Psi(\phi_0(x))=\int_{\phi(t_E=-\infty)}^{\phi(t_E=0, x)=\phi_0(x)}D\phi\ e^{-S(\phi)},
\eea
where $t_E$ is the Euclidean time, $x$ is the Euclidean space and $S$ is action. Thus, a density matrix is given by
\bea
(\rho_V)_{\phi_+\phi_-}&=&\Psi(\phi_-)\Psi^*(\phi_+)
\nn\\
&=&(Z_1)^{-1}\int_{t_E=-\infty}^{t_E=\infty}D\phi\ e^{-S(\phi)}\ \Pi_{x\in V}\
\delta \bigg(\phi(0^+, x)-\phi_+(x)\bigg)\delta\bigg(\phi(0^-, x)-\phi_-(x)\bigg),
\nn\\
\eea
where a choosing of $Z_1$ is to let $\mbox{Tr}\rho_V=1$. Our decomposition of space is $V=A+B$ and $B=C\cup D$, where region $A$ is entangling surface and region $B$ is bulk region. We want to get a reduced density matrix in region $C$.  Now we remove some operators in region $A$ to let center live in region $A$ to discuss center issue in the Lagrangian formulation.  To obtain a reduced density matrix with a center, we let $\phi_+=\phi^A_+\ \oplus\ \phi^B_+$ and $\phi_-=\phi^A_-\ \oplus\ \phi^B_-$, where we denote $A$ be entangling surface and denote $B$ be bulk region. Then we set $\phi_+=\phi_-$ in the $A$ region (due to center in region $A$) and $D$ region (due to integrating out the field in region $D$). The center effects appear in the entangling surface ($A$ region) so it is equivalent to setting a boundary condition. We can explicitly work this procedure in free theory. Now we decompose our fields into classical part and quantum part. The center comes from classical state because $\lbrack\phi, \phi\rbrack=0$. This implies that we do not have quantum fluctuation on the entangling surface. The quantum part vanishes on the entangling surface so we only have quantum fluctuation on the bulk. If we decompose our fields arbitrary, we possibly have singularity in path integral. In order to avoid this problem, we choose classical background and do quantum fluctuation around classical background. The classical background leads on-shell action vanishes on the bulk in free theory when we do partial integration by part. We only have boundary on-shell action in free theory \cite{Donnelly:2015hxa, Donnelly:2014fua}. The boundary field also decouples from the bulk field on the bulk. Finally, we sum over all classical configuration and quantum fluctuation, then we get a reduced density matrix. In interaction theory, the boundary field will couple to bulk field and the on-shell action may not vanish on the bulk. This is also consistent with the Hamitonian formulation \cite{Casini:2013rba}. The boundary term should correspond to classical Shannon entropy from our analysis. It is useful to check whether the decomposition is the tensor product decomposition in the Hilbert space from classical solutions on the boundary.

\subsubsection{Replica Trick and Conical Method}
To compute the entanglement entropy, we use the replica trick and conical method to know the relation between the entanglement entropy and partition function. The entanglement entropy in the replica trick can be rewritten as
\bea
S_A=\lim_{n\rightarrow 1}\frac{\mbox{Tr}(\rho_A^n)-1}{1-n}=-\frac{\partial}{\partial n}\mbox{Tr}\rho_A^n\bigg|_{n=1}.
\eea
To compute $\mbox{Tr}\rho_A^n$, we do $n$ copies
\bea
(\rho_A)_{\phi_{1+}\phi_{1-}}(\rho_A)_{\phi_{2+}\phi_{2-}}\cdots (\rho_A)_{\phi_{n+}\phi_{n-}}
\eea
with $\phi_{i-}=\phi_{(i+1)+}$. Then path integral representation for $\mbox{Tr}\rho_A^n$ on $n$-sheet space is 
\bea
\mbox{Tr}\rho_A^n=(Z_1)^{-n}\int D\phi\ e^{-S(\phi)}.
\eea
The entanglement entropy in the conical method is given by
\bea
S_A=\bigg(1-\beta\frac{\partial}{\partial\beta}\bigg)\ln Z(\beta)\bigg|_{\beta=2\pi}
\eea
with $\beta=2\pi n$ and $Z(\beta)=Z(2\pi n)=Z(2\pi)^n\mbox{Tr}\rho_A^n$. Now we show that the conical method can also give the entanglement entropy as
\bea
\bigg(1-\beta\frac{\partial}{\partial\beta}\bigg)\ln Z(\beta)\bigg|_{\beta=2\pi}
&=&\bigg(1-n\frac{\partial}{\partial n}\bigg)\ln Z(2\pi n)\bigg|_{n=1}
\nn\\
&=&\ln \bigg(Z(2\pi)\bigg)-n\ln \bigg( Z(2\pi)\bigg)-\frac{1}{\mbox{Tr}\rho^n}\frac{\partial}{\partial n}\mbox{Tr}\rho^n\bigg|_{n=1}
\nn\\
&=&-\mbox{Tr}\bigg(\rho\ln \rho\bigg).
\eea
The replica trick and conical method are equivalent. The entanglement entropy for the one-form abelian Yang-Mills gauge theory is computed by using the conical method in \cite{Donnelly:2015hxa, Donnelly:2014fua}. 

\subsection{The Einstein Gravity Theory}
We discuss the entanglement entropy in the Einstein gravity theory and discuss the holographic entanglement entropy.

\subsubsection{Entanglement Entropy in the Einstein Gravity Theory}
The entanglement entropy in gauge theories suffers from gauge symmetry is hard to define the entanglement entropy in the trivial choice and continuum limit. The Einstein gravity theory also has diffemorphism so we possibly face the similar situation. Due to the Einstein gravity theory is gauge invariant by performing partial integration by part, the entanglement entropy in the Einstein gravity theory should be more subtle than gauge theory. We use the saddle point approximation to compute the entanglement entropy. The cosmological constant term will contribute the bulk on-shell action. We also ignore all quantum fluctuation. The details of the entanglement entropy for the Einstein gravity theory is given in Appendix \ref{app5} \cite{Ryu:2006bv, Lewkowycz:2013nqa, Fursaev:1995ef}. From Appendix \ref{app5}, we use the saddle point approximation to obtain the entanglement entropy $A_{q-2}/(4G)$ which only comes from entangling surface \cite{Ryu:2006bv, Lewkowycz:2013nqa, Fursaev:1995ef}, where $A_{q-2}$ is codimension two surface. This means that the entanglement entropy is not a trivial choice and the leading order computation gives codimension two surface term. This result is expected because gravity theory can be constructed from gauge formulation. Gravity theory has many similar properties with gauge theory so gravity theory should have the same problem in the entanglement entropy of the trivial choice on continuous space. This motivates us to study more about the entanglement entropy with the non-trivial centers. Otherwise, the entanglement entropy in gravity theory is hard to define.

\subsubsection{Comments in the Holographic Entanglement Entropy}
We discuss application of the holograph principle in the entanglement entropy. The holograph principle is motivated from string theory or $AdS_5/ CFT_4$ in some limits \cite{Maldacena:1997re}. The computation related to $AdS_5$ metric is given by Appendix \ref{app6}. When we take $N\rightarrow \infty$ and $g_sN\gg 1$, where  $N$ is rank of the gauge group and $g_s$ is string coupling constant, the metric will approach to flat metric. Hence, we can apply our previous result (the entanglement entropy in the Einstein gravity theory) to $AdS_5$. The belief of the $AdS/CFT$ correspondence comes from the string interpretation. In low-energy effective theory, we can take limit to see some clues for $AdS_5/CFT_4$ from the multiple $D3$-branes solution  \cite{Maldacena:1997re}. Now we discuss the limit \cite{Maldacena:1997re} for the entanglement entropy. We take limit as
\bea
r\rightarrow 0, \qquad l_s\rightarrow 0 , \qquad N\rightarrow\infty, \qquad g_s\rightarrow 0, \qquad g_sN \gg 1
\eea
with $l_s^2/r$ and $g_sN$ fixed, where $r$ is distance that parallel D3-branes separated,  $l_s$ is string length. When we take $r\rightarrow 0$ and $l_s\rightarrow 0$ with $l_s^2/r$ fixed, we can obtain $AdS_5$ metric from multiple D3-branes solutions \cite{Maldacena:1997re}. Finally we want to make our gravity theory and gauge theory can be computed in a suitable limit so we take $N\rightarrow \infty$, $g_s\rightarrow 0$ and $g_s N\gg 1$ with fixed $g_sN$. These conditions are compatible with our computation of the entanglement entropy. We can study the holograph principle and obtain codimension two surface. The holograph principle is useful in the application of knowing behavior of strongly coupled gauge theory. Our computation of the entanglement entropy in leading order of the Einstein gravity is proportional to the codimension two surface. We can also do perturbation analysis with periodic fields to relate the codimension two surface to minimum surface \cite{Ryu:2006bv, Lewkowycz:2013nqa}. Our analysis should define the reduced density matrix in the holographic entanglement entropy.

\section{Strong Coupling Expansion of the $SU(N)$ Lattice Gauge Theory}
\label{6}
We use the Haimitonian formulation to compute the entanglement entropy of the $SU(N)$ Yang-Mills gauge theory in the fundamental representation on lattice in strong coupling limit. The strong coupling limit is hard to compute from field techniques. On lattice, we use the strong coupling expansion to compute. Due to difficulties of defining the entanglement entropy, we extend the Hilbert space to obtain the gauge invariant entanglement entropy. This extended space of the entanglement entropy is the electric choice of the entanglement entropy. We introduce lattice Hamitonian formulation of the $SU(N)$ Yang-Mills gauge theory in the fundamental representation in Appendix \ref{app7}, and review the extended lattice model in Appendix \ref{app8}.

\subsection{Equivalence between the Electric Choice and the Extended Lattice Model}
The extended lattice \cite{Buividovich:2008gq, Donnelly:2011hn} is a useful way to compute the entanglement entropy in the lattice gauge theory. The main idea is to add more degrees of freedom on the entangling surface to define a gauge invariant entanglement entropy. A nature question should arise: What choice for the entanglement entropy in the extend Hilbert space? Now we show that the extend lattice model can give us an electric choice of the entanglement entropy in the lattice gauge theory with \emph{finite} lattice spacing. We first prove that expectation values of all original operators do not change. The operators in the interior $V^i$ do not change from \eqref{embedding}. For the operators on the entangling surface, this follows from
\bea
\lag\hat{L}^{l_{\partial V}}_{g}\rag &=&\int (\Pi_{l\in L^\prime} dU_l) \,\psi^\prime(U_{l_1},\cdots,U_{l_{\partial V}},U_{l_{\partial \bar{V}}},\cdots,U_{l_n})^*\, \hat{L}^{l_{\partial V}}_{g}\, \psi^\prime(U_{l_1},\cdots,U_{l_{\partial V}},U_{l_{\partial \bar{V}}},\cdots,U_{l_n})
\nn\\
&=&\int (\Pi_{l\in  L^\prime} dU_l) \,\psi^\prime(U_{l_1},\cdots,U_{l_{\partial V}},U_{l_{\partial \bar{V}}},\cdots,U_{l_n})^* \psi^\prime(U_{l_1},\cdots,g U_{l_{\partial V}},U_{l_{\partial \bar{V}}},\cdots,U_{l_n})
\nn\\
&=&\int (\Pi_{l\in L^\prime} dU_l) \,\psi(U_{l_1},\cdots,U_{l_{\partial V}}U_{l_{\partial \bar{V}}},\cdots,U_{l_n})^* \psi(U_{l_1},\cdots,g U_{l_{\partial V}}U_{l_{\partial \bar{V}}},\cdots,U_{l_n})
\nn\\
&=&\int (\Pi_{l\in L} dU_l) \,\psi(U_{l_1},\cdots.,U_{l_{\partial}},\cdots,U_{l_n})^* \hat{L}_g^{l_\partial}\psi(U_{l_1},\cdots, U_{l_{\partial}},\cdots,U_{l_n})\,,
\eea
where we used 
\bea
(\hat{L}_g^{l} \psi)\lbrack U_1,\cdots,U_n\rbrack =\psi\lbrack U_1,\cdots,g U_{l},\cdots,U_n\rbrack\,, \qquad \hat{L_g^l}\equiv \exp\big(i\omega^b_lE_l^b\big), \qquad g\equiv \exp\big(i\omega_l^aT_l^a\big).
\nn\\
\eea
Finally, we show that a reduced density matrix is given by
\bea
\rho_V[U_{V^i},U_{\partial V},U_{V^i}^\prime,U_{\partial V}^\prime]=\int (\Pi_{l_{\bar{V}}\in  L_{\bar{V}}} dU_{l_{ \bar{V}}})\,\psi\lbrack U_{V^i},U_{\bar{V}^i},U_{\partial V}U_{\partial \bar{V}} \rbrack\psi^*\lbrack U_{V^i}^\prime,U_{\bar{V}^i},U_{\partial V}^\prime U_{\partial \bar{V}}\rbrack\,\label{episte},
\nn\\
\eea
which commutes with the link operator $\hat{L}_g^{\partial V}$ on entangling surface. The proof is given by
\bea
\hat{L}_g^{l_{\partial V}}\rho_V\lbrack U_{V^i},U_{\partial V},U_{V^i}^\prime,U_{\partial V}^\prime\rbrack&=&\int (\Pi_{l\in L_{\bar{V}}} dU_{l})
\,\psi\lbrack U_{V^i},U_{\bar{V}^i},g U_{\partial V}U_{\partial \bar{V}} \rbrack
\psi^*\lbrack U_{V^i}^\prime,U_{\bar{V}^i},U_{\partial V}^\prime U_{\partial \bar{V}}\rbrack
\nn\\
&=& \int (\Pi_{l\in L_{\bar{V}}} dU_{l}) \, \psi\lbrack U_{V^i},U_{\bar{V}^i}, U_{\partial V}U_{\partial \bar{V}} \rbrack\psi^*\lbrack U_{V^i}^\prime, U_{\bar{V}^i},g^{-1} U_{\partial V}^\prime U_{\partial \bar{V}}\rbrack 
\nn\\
&=& \rho_V\lbrack U_{V^i},U_{\partial V},U_{V^i}^\prime,U_{\partial V}^\prime\rbrack\hat{L}_g^{l_{\partial V}}\,.
\label{episte1} 
\nn\\
\eea
All local operators in the extended lattice model \cite{Buividovich:2008gq, Donnelly:2011hn} does not change from extending the Hilbert space, and the reduced density matrix of the extended lattice model commutes with the link operator on entangling surface. This establishes the equivalence of the entanglement entropy between the extended lattice model \cite{Buividovich:2008gq, Donnelly:2011hn} and the electric choice. When we take continuum limit, the result should give the electric choice of the entanglement entropy without enlarging the Hilbert space.

\subsection{Strong Coupling Expansion}
The Hamiltonian for the $SU(N)$ Yang-Mills gauge theory in the fundamental representation on a lattice of spacetime dimensions
$D > 2$ is given by
\begin{equation}
H_{\mbox{LYMF}} = \frac{g^2}{2}\sum_l E_l^a E_l^a - \frac{1}{g^2}\sum_\sq\left(\Tr U_\sq + \Tr U_\sq^\hc\right) \,.
\end{equation}
We denote the Lie algebra indices as $a-h$.

It is more convenient to work with a dimensionless quantity
\begin{equation}
W_{\mbox{LYMF}}  = \frac{2}{g^2}H_{\mbox{LYMF}} = W_E + c\,\lam\,W_B \,,
\end{equation}
where
\begin{equation}
W_E = \sum_l E_l^a E_l^a \,, \qquad 
W_B = \sum_\sq\left(\Tr U_\sq + \Tr U_\sq^\hc\right) \,, 
\qquad c\,\lam = -\frac{2}{g^4} \,.
\end{equation}
In strong coupling limit $g \gg 1$ and $\lam \ra 0$, ground state can be calculated by treating $W_B$ as a perturbation with unperturbed eigenstates of $W_E$.

The ground state of $W$ can now be calculated using standard perturbation theory. To the second order, this is given by
\begin{align}
|\Omega\rag &= 
\left(1 - \half\lam^2 N_\sq\right)|0\rag + \lam\sum_\sq|\sq\rag + \cdots+ O(\lam^3) \,,
\end{align}
where we used that the unperturbed single-loop eigenstate energy is
\begin{equation}
E_\sq^{(0)} = \lag\sq|W_E|\sq\rag = 4C_2(N)\lag\sq|\sq\rag \,,
\end{equation}
with $C_2(N) = \frac{N^2 - 1}{2N}$ is the quadratic Casimir of $SU(N)$ in the fundamental representation. We also set $c = -2C_2(N)$ for simplification and used the commutation relations between the electric field and the plaquette field, which is given by
\begin{equation}\label{eq:commrel}
[E_l^a\,,U_{l'}] = \delta_{ll'}T^a U_l \,, \qquad [E_l^a\,,U_{l'}^\hc] = -\delta_{ll'}T^a U_l^\hc \,,
\end{equation}
for the unperturbed single-loop eigenstate energy. In the expression for $|\Omega\rag$ above, $N_\sq$ is the total number of plaquettes on the lattice. Although $\cdots$ in the $|\Omega\rag$ is at order of $\lambda^2$, they will be irrelevant in the entanglement entropy calculations below.  The ground state at the zeroth order is the state where electric field vanishes
\begin{equation}
E_l^a E_l^a|0\rag = 0  \,
\end{equation}
for each link $l$,
and a operation of $\Tr U_\sq$ acting on the ground state is to excite a loop of single flux with positive orientation (negative orientation if $\Tr U_\sq^\hc$)
\begin{equation}
\Tr U_\sq|0\rag = |\sq\rag \,. 
\end{equation}

The entanglement entropy in the extended lattice model is given by~\cite{Donnelly:2011hn}
\begin{equation}\label{eq:EEetl}
S(\rho_A) = H_C(p(R_\pd)) + \sum_{l \in L_{\pd A}}\lag\ln\mathrm{dim}(r_l)\rag + \lag S(\rho_A(R_\pd))\rag \,,
\end{equation}
where $p(R_\pd)$ is the probability distribution of irreducible representations on the boundary, where
$R_\pd = \{r_l: l \in L_\pd\}$ ($L_\pd = L_{\pd A} \cup L_{\pd B}$) is the set of oriented links crossing the boundary, and $\rho_A(R_\pd)$ is the reduced density matrix associated with $R_\pd$. 

At $O(\lam^2)$, only the trivial state and the single-plaquette states will contribute to $p(R_\pd)$. Let $n_A$ be the number of boundary links in region $A$. The number of single-plaquettes intersecting the boundary is
\begin{equation}
N_\sq(\pd) = n_A(D-2) \,
\end{equation}
since there are $(D-2)$ degrees of freedom of plaquette that we can take. Then there are $N_\sq(\pd)$ different possible sets of (nontrivial) $R_\pd$ with probability $\lam^2$, and the probability of no intersection with the boundary (trivial $R_\pd$) is $1 - N_\sq(R_\pd)\lam^2$. The classical Shannon entropy is thus
\begin{align}
H_C(p(R_\pd)) &= -N_\sq(\pd)(\lam^2\ln\lam^2) - (1 - N_\sq(R_\pd)\lam^2)\ln(1 - N_\sq(R_\pd)\lam^2) \notag \\
&= n_A(D-2)\lam^2(-\ln\lam^2 + 1) + O(\lam^3) \,.
\end{align}
Next, each single-plaquette loop intersects the boundary in two links in the $SU(N)$ fundamental representation $N$, the second term in \eqref{eq:EEetl} is $2n_A(D-2)\lam^2\ln N$. Lastly, to calculate $\lag S(\rho_A(R_\pd))\rag$ up to the order of $O(\lam^2)$, $\rho_A(R_\pd)$ is composed of pure states, since only the trivial and the single-plaquette states contribute at this order. In the single-plaquette case, only $\rho_A(R_\pd) = \lam^2|\sq\rag\lag\sq|$ is possible, which is a pure state. We also find a pure state when considering the trivial state case. Thus, $\lag S(\rho_A(R_\pd))\rag = 0$, and the entanglement entropy \cite{Radicevic:2015sza} is
\begin{equation}
S_{\mbox{EE}} =  n_A(D-2)\lam^2(-\ln\lam^2 + 1 + 2\ln N) + O(\lam^3) \,.
\end{equation}
The strong coupling expansion of the entanglement entropy in the Lagrangian formulation is discussed in \cite{Chen:2015kfa}.

Now if the theory is a $U(1)$ theory, the entanglement entropy would just be the classical Shannon term, since the abelian gauge theory has only one-dimensional representations so the second term in  \eqref{eq:EEetl} vanishes. 

Our computation shows that the most important contribution is $\ln \lambda^2$ in the classical Shannon term. The classical Shannon term comes from the center or ambiguity. This result reflects that the choice of the ambiguity becomes important in strong coupling limit. But the choice of center should come from the entangling surface. The degrees of freedom on the bulk is larger than the degrees of freedom on the entangling surface. Even for the strong coupling limit, the reason that a choice of ambiguity becomes dominant is still unclear. This term is similar with codimensional two surface term in gravity theory. They can be gotten from the saddle point approximation. But we expect that this term will be canceled in the mutual information on continuous space. 

When considering the large $N$ limit, $\ln N$ will be compatible with the classical Shannon term due to $\lambda\sim \frac{1}{N}$. In infinity strong coupling limit, the entanglement entropy will vanish. In strong coupling region, we expect color confinement. The color confinement phenomenology gives  singlet state so the entanglement entropy will vanish. Our result gives a consistent result with the color confinement. 

The entanglement entropy of gauge theory is also computable from the Monte-Carlo simulation \cite{Buividovich:2008kq, Nakagawa:2009jk, Nakagawa:2011su}. Our results shows that the entanglement entropy is proportional to spatial area terms, which are also confirmed from the lattice simulation.

The above conclusions and observations are based on the strong coupling expansion. But we remind that the strong coupling limit has a drawback in the continuum limit. Our conclusion may not give a correct understanding in the strong coupling region. This situation is similar with the confinement of the QCD. We also use the strong coupling expansion to obtain the confinement. Although we possibly lose continuum limit in the strong coupling limit, we still believe that the strong coupling expansion gives us some reliable properties.  However, some results of the entanglement entropy are still puzzle. We need to use other ways or toy models to get consistent understanding with the strong coupling expansion. We leave these interesting studies to future.

\section{Conclusion}
\label{7}
We discuss the entanglement entropy with center from various ways. The entanglement information in quantum field theory can be understood from algebra. The algebraic approach is generic and rigorous. In local quantum field theory, the Von-Neumann algebra does not lose generality with complete discussion on properties of the entanglement information. The important properties of the entanglement information come from the partial trace operators, and strong subadditivity or other inequalities with information meaning. In non-trivial choice, these studies are not clearly understood. We use a mathematical point of view to discuss these properties. In our analysis, the strong subadditivity may not be satisfied from generic centers. In our discussion, we can understand how to choose the decomposition of the Hilbert space to get the strong subadditivity. In these cases, the partial trace operator will also give the entanglement entropy which contains information. We use the Hamitonian formulation to compute the entanglement entropy to study global symmetry between different centers. We find duality structure to show continuous entanglement entropy in some theories. It is interesting to use this method to find duality structure in centers or find possibility to classify entanglement entropy. The Hamitonian approach is useful to study properties of the entanglement entropy, but it is hard to get exact solutions. We propose the Lagrangian formulation from the Hamitonian formulation. The Lagrangian formulation is useful due to that we do not need to worry how to define the partial trace operator during computation of the entanglement entropy. We also use the Lagrangian formulation to discuss decomposition of the Hilbert space in the Einstein gravity theory. Our results show that codimension two surface term comes from a non-trivial choice, and the surface term is also the leading order result. This reflects that gravity theory is also hard to define the entanglement entropy in the trivial choice with gauge symmetry as in gauge theory. The reason is due to that the gauge invariant gravity theory also relies on boundary conditions. This result is not surprising from the similarity between gauge and gravity theories. This example also sheds light on center issues for the decomposition of the Hilbert space in the holographic entanglement entropy, and validity of the holograph principle in the entanglement entropy. When considering the gauge theory, gauge symmetry will be broken if you naively choose an entangling surface. In the case of gravity theory, gauge invariance relies on the boundary condition so we also suffer from the same problem as in the gauge theory. Because the gravity theory can be rewritten from the gauge formulation, this is expected. Two major theories in high energy theory do not have a suitable definition for the entanglement entropy in the trivial choice, then we should have motivation to consider non-trivial centers in the entanglement entropy. Finally, we consider the extended lattice model to compute the entanglement entropy in the $SU(N)$ lattice Yang-Mills gauge theory in the fundamental representation. The motivation is the behavior of the entanglement entropy in the strong coupling region. The computation in strong coupling region is hard to perform, but it is easier to compute from the strong coupling expansion in the lattice method. The strong coupling expansion is not a fully self-contained way because the lattice gauge theory possibly not have continuum limit in strong coupling limit. The confinement issue also has the same problem, and we believe that the strong coupling expansion should give some phenomenological understanding. The strong coupling expansion possibly still gives reliable behaviors in the entanglement entropy. However, we need to offer more consistent understanding for our puzzles of the lattice gauge theory in strong coupling region.

The entanglement entropy with center is defined on choosing a suitable basis. The entanglement entropy in different regions may not be detected from same basis. The proof of the strong subadditivity suffers from this problem. We point out this issue in our paper and think that this problem may be a key issue in physical interpretation of the entanglement entropy. The modification of the strong subadditivity \cite{Casini:2014aia} or the partial trace operator to define the entanglement entropy may be a way to solve this problem.

When considering centers in the entanglement entropy, we need to remove some operators from an entangling surface. We interpret that different choices of centers come from different observations (or different partial trace operators) in the entanglement entropy. It is nature to know that different centers have their respective entanglement information. But the entanglement information with non-trivial centers is expected to be the same as the entanglement information with the trivial center. Let us point out two problems related to the classical Shannon entropy. The entanglement entropy with the non-trivial centers is the combination of the quantum entropy and classical Shannon entropy. The classical Shannon entropy comes from a probability distribution of centers. Hence, we can interpret that this term gives us an additional classical entanglement information on an entangling surface. We already found universal contributions from the classical Shannon entropy \cite{Kitaev:2005dm, Levin:2006zz, Pretko:2015zva}. But we have two types classical Shannon entropy. The first type of the classical Shannon entropy comes from quantum entropy, which is still tensor product decomposition. The second type of the classical Shannon entropy comes from non-tensor product decomposition. Although they have a same form as the classical Shannon entropy, but their meaning should be different. The entanglement information with centers is not totally clear now. Hence, we need to understand more about universal contributions from the classical Shannon entropy. The Lagrangian formulation should be a useful tool to understand more. The second problem is how to observe universal term in the classical Shannon entropy or on the entangling surface. The mutual information possibly counts the bulk degrees of freedom without concerning the degrees of freedom on the entangling surface. How to find a combination of the entanglement entropy to find the entanglement information related to entangling surface is an interesting direction.

The entanglement entropy in the Chern-Simons, supersymmetric or other gauge theories are already computed by the replica trick. Their results are gauge invariant. But one problem is whether the gauge invariance can guarantee what we compute is the entanglement entropy. This issue does not have serious computation to check or what decomposition of the Hilbert space corresponds to. We leave this interesting work to future.

Many properties of the entanglement entropy can be determined from algebra without explicit computing. Some interesting problems related to quantum gravity are what the Hilbert space can give the ultraviolet complete quantum gravity. The ultraviolet complete quantum gravity should not suffer any problems from centers, then the non-commutative geometry is a candidate because the non-commutative structures can avoid center ambiguities when we decompose the Hilbert space. The other approach is to define our quantum gravity on the discrete space. Then the entanglement entropy will not suffer from the problem of regulators. But this approach is hard to find theoretical evidences in low-energy regions. Some interesting applications are reading strong coupling information from the entanglement entropy. In strong coupling region, we have mysterious phase structure. The entanglement entropy does not vanish at zero temperature so it may be a useful order parameter to classify the phase structure. Before we work on this problem in QCD, the idea of the classification can be tested from some simple toy models.

The entanglement entropy is hard to compute in field theory so the holograph method for the first understanding is necessary. Our paper should shed the light on understanding validity of the holograph method on entangling surface or relating the center to the holographic entanglement entropy or conformal field theory from the decomposition of the Hilbert space. The other interesting direction should be the algebraic approach. The computation of the entanglement entropy is related to partition function, and can be determined from the simple two-point functions in free theory \cite{Casini:2009sr}. Two-point functions in free theory may give us a hint or first step to understand how to construct the entanglement entropy from an algebraic point of view. Algebraic structure of the entanglement entropy possibly gives us more rigorous quantum properties without constructing the Lagrangian and Hamitonian densities.

\acknowledgments

The author would like to thank H. Casini, Jun-Kai Ho, Xing Huang, Ling-Yan Hung, J. A. Rosabal, Tadashi Takayanagi and Chien-Hsun Wang for their useful discussion. Especially, the author would like to thank Nan-Peng Ma for his suggestion and encouragement, and also thank Mu-Sheng Wu for his contribution in the strong coupling expansion part.

\appendix

\section{Information}
\label{app1}
Information is an abstract concept, but we expect that this quantity measure number of states precisely. A suitable definition of information should be unique under conditions or constraints. Let us show how to define information uniquely. We use how surprise to define information. Now we use four axioms to define information as
\begin{enumerate}[label=\textbf{Axiom\arabic*.}]
  \item \label{item:A1} $Sur(1)=0$.

$Sur(p)$ is a function of probability $p$, and this function defines how surprise that we received from one event. If the probability of one event occurs equals one, we should not surprise. This is the reason for the first axiom.
 \item \label{item:A2} If $p< q$, $Sur(p)> Sur(q)$. 

More probability on a event should decrease your surprise. This is why we have the second axiom.

 \item \label{item:A3} $Sur(p)$ is a continuous function of $p$.

We believe that information that received should be continuous.  
 \item \label{item:A4} $Sur(pq)=Sur(p)+Sur(q).$ 

If two variables are independent, information has the additive property.
\end{enumerate}

\begin{theorem}
If $Sur(p)$ satisfies from \ref{item:A1} to \ref{item:A4}, then $Sur(p)=-C\log_2 p$, where $C$ is an arbitrary positive integer.
\end{theorem}
\begin{proof}[Proof]
 We first use \ref{item:A4} to obtain
\bea
Sur(p^2)=Sur(p)+Sur(p)=2Sur(p),
\eea
then we do induction to get
\bea
Sur(p^m)=mSur(p).
\eea
For all integers $n$, $Sur(p)=Sur(p^{\frac{1}{n}}\cdots p^{\frac{1}{n}})=nSur(p^{\frac{1}{n}})$. Hence, we have
\bea
Sur(p^{\frac{1}{n}})=\frac{1}{n}Sur(p).
\eea
This is easier to do generalization to get
\bea
Sur(p^{\frac{m}{n}})=mSur(p^{\frac{1}{n}})=\frac{m}{n}Sur(p).
\eea
This is equivalent to
\bea
Sur(p^x)=xSur(p),
\eea
where $x$ is a positive rational number.
Due to the continuity condition in (\ref{item:A3}), $x$ can be extended to non-negative regions. Let $x=-\log_2p$ for $0<p\le 1$, then 
\bea
Sur(p)=Sur(2^{-x})=xSur(\frac{1}{2})=-C\log_2 p,
\eea
where $C=Sur(\frac{1}{2})>Sur(1)=0$.
\end{proof}

This theorem shows a unique form to define information from how surprise. A quantity related to information is defined by expectation value of $-\log_2 p$, where $p$ is probability. 
 Due to that the quantum entropy has $S_Q(\rho)=S_Q(U^{\dagger}\rho U)$, where $U^{\dagger}U=1$, we will find that a similar form between the classical Shannon entropy and quantum entropy so the quantum entropy is a natural generalization from the classical Shannon entropy. The entanglement entropy with the non-trivial center is the combination of quantum entropy and classical Shannon entropy. The classical Shannon entropy is a unique expression, and the quantum entropy is generalized from the classical Shannon entropy with a compact form. We possibly not have other choices to define the entanglement entropy with gauge symmetry.

\section{Review of the Von-Neumann Algebra}
\label{app2}
In local quantum field theory, discussion of the Von-Neumann algebra in the Hilbert space are complete without losing generality. This also leads us to consider the generic cases of the entanglement entropy \cite{Casini:2013rba}. We review useful theorems \cite{von} in the Von-Neumann algebra that we will use in the entanglement entropy. 

\subsection{Definition of the Von-Neumann Algebra}
We denote $H$ as a complex Hilbert space, and $L(H)$ as continuous linear operators from $H$ to $H$. The commutant of $M$ is $M^{\prime}$. This means that all elements of $L(H)$ which commute with all elements of $M$. The bicommutant is defined as $(M^{\prime})^{\prime}=M^{\prime\prime}$. From the bicommutant, it is obvious to get $M\subseteq M^{\prime\prime}$. If $M\subseteq N$, it also implies $M^{\prime}\supseteq N^{\prime}$ and $M^{\prime\prime}\subseteq N^{\prime\prime}$. It is easy to deduce $M^{\prime}\supseteq (M^{\prime\prime})^{\prime}=M^{\prime\prime\prime}$, and $M^{\prime}\subseteq (M^{\prime})^{\prime\prime}=M^{\prime\prime\prime}$ (We replace $M$ by $M^{\prime}$.) . We can find a general result as
\bea
M^{\prime}=M^{\prime\prime\prime}=\cdots=M^{(2n^{\prime}-1)}=\cdots, \qquad M\subseteq M^{\prime\prime}=M^{(4)}=\cdots=M^{(2n^{\prime})}=\cdots,
\nn\\
\eea
where $2n^{\prime}-1$ factor in $M^{(2n^{\prime}-1)}$ means numbers of prime, and we also denote numbers of prime as from $n^{\prime}$ to $z^{\prime}$, ranging from one to infinity. In $L(H)$, we have an adjoint operation. If $S\in L(H)$, we will denote the adjoint of $S$ as $S^{\dagger}$. Hence, we have 
\bea
(S+T)^{\dagger}=S^{\dagger}+T^{\dagger}, \qquad (\lambda S)^{\dagger}=\lambda^*S^{\dagger}, \qquad (ST)^{\dagger}=T^{\dagger}S^{\dagger}, \qquad S^{\dagger\dagger}=S,
\eea
where $\lambda$ is a complex number, and $\lambda^*$ is complex conjugate of $\lambda$. $L(H)$ can be a *-algebra (or involutive algebra). Each algebra in $L(H)$ is stable under the adjoint operation.
\begin{definition}
The Von-Neumann algebra is a *-subalgebra $A$ in $L(H)$ which satisfies $A=A^{\prime\prime}$ in $H$.
\end{definition}
The algebra $L(H)$ is the Von-Neumann algebra which always contains scalar operator. The collection of the scalar operator are denoted as $C_H$.

Let $M$ be an adjoint stable subset of $L(H)$. The set $M^{\prime}$ and $M^{\prime\prime}$ are the Von-Neumann algebras. If $A$ is the Von-Neumann algebra with $M\subseteq A$, we should have $M^{\prime\prime}\subseteq A^{\prime\prime}=A$. Therefore, $M^{\prime\prime}$ is the smallest Von-Neumann algebra which contains $M$. If $M$ is any subset of $L(H)$, and $N=M\cup M^{\dagger}$ ($M^{\dagger}$ is an image of $M$ under the adjoint operation.) The Von-Neumann algebra containing $M$ are those containing $N$. Thus, $N^{\prime\prime}$ is the smallest Von-Neumann algebra which contains $M$, and $N^{\prime\prime}$ is called the Von-Neumann algebra generated by $M$.  We also denote the closed linear subspace of $H$ generated by $Tx$ ($T\in A$, $x\in M$) as $X_M^A$.
\begin{definition}
A factor is the Von-Neumann algebra whose center only contains scalar operator.
\end{definition}
\begin{definition}
The Von-Neumann algebra $A$ is said to be $\sigma$-finite if every family of non-zero pairwise orthogonal projections of $A$ is countable. In a separable Hilbert sapce, every Von-Neumann algebra is $\sigma$-finite.
\end{definition}
\begin{definition}
The Borel space is a set endowed with a set $B$ of subsets of $E$ which has following properties: If $B$ is closed under countable unions and taking of complements (and under countable intersections), elements in $B$ are called the Borel sets of $E$.
\end{definition}
\begin{definition}
For each $\xi\in z$, where $z$ is a Borel space, let $A(\xi)$ be the Von-Neumann algebra in $H(\xi)$. The mapping $\xi\rightarrow A(\xi)$ is called a field of the Von-Neumann algebra over $z$.
\end{definition}
\begin{definition}
A field of the Von-Neumann algebra $\xi\rightarrow A(\xi)$ over $z$ is said to be measurable if there are sequences $\xi\rightarrow T_1(\xi),\ \xi\rightarrow T_2(\xi),\ \cdots$ of measurable fields of operators almost everywhere. $A(\xi)$ is the Von-Neumann algebra generated by $T_1(\xi)$, $T_2(\xi)$, $\cdots$.
\end{definition}
\begin{definition}
The Von-Neumann algebra $A$ in $H$ is called decomposable if it is defined by a measurable field $\xi\rightarrow A(\xi)$ of the Von-Neumann algebras. We write
\bea
A=\int^{\bigoplus} A(\xi)\ d\nu(\xi).
\eea
 If all $A(\xi)$ are scalar operators, the Von-Neumann algebra is called diagonalizable.
\end{definition}

\subsection{Topology in the Von-Neumann Algebra}
We will define topology that we will use in the Von-Neumann algebra. These topologies are essential to show useful theorems in the Von-Neumann algebra.

Let $x\in H$. The map $T\rightarrow\ || Tx||$ is a seminorm ($|| v||=0$ is equivalent to $v=0$) in $L(H)$. A collection of all these seminorms determines the Hausdorff locally convex topology in $L(H)$ called topology of strong pointwise convergence or strong topology.

Let $x, y\in H$. Function $T\rightarrow\ | \lag Tx, y\rag|$, where $\lag A, B\rag$ is inner product space between $A$ and $B$, is a seminorm in $L(H)$. A collection of all these seminorms defines the Hausdorff locally convex topology in $L(H)$ called topology of weak pointwise convergence or weak topology.

Let $(x_1, x_2, \cdots)$ be a sequence of elements in $H$ with $|| x_1||^2+|| x_2||^2+\cdots \ <\ \infty$. For each $T\in L(H)$, we have
\bea
|| Tx_1+Tx_2+\cdots ||^2\ \le\ ||T||^2\bigg( ||x_1||^2+|| x_2||^2+\cdots\bigg)    \ <\ \infty
\eea
and function $T\rightarrow\ \bigg(||Tx_1||^2+||Tx_2||^2+\cdots\bigg)^{1/2}$ is a seminorm in $L(H)$. A collection of all these seminorms defines the Hausdorff locally convex topology in $L(H)$ called ultra-strong topology.

Let $(x_1, x_2, \cdots)$, $(y_1, y_2, \cdots)$ be two sequences  of elements in $H$ with
\bea
|| x_1||^2+||x_2||^2+\cdots \ <\ \infty, \qquad || y_1||^2+||y_2||^2+\cdots\ <\ \infty.
\eea 
For each $T\in L(H)$, we have
\bea
&&|\lag Tx_1, y_1\rag|\ \le\ || T x_1||\ || y_1||+|| T x_2||\ || y_2||+\cdots \le\ ||T||\bigg( ||x_1||\ || y_1||+||x_2||\ || y_2||+\cdots\bigg)
\nn\\
&& \le\ || T||\bigg(|| x_1||^2+|| x_2||^2+\cdots\bigg)^{\frac{1}{2}}\bigg(|| y_1||^2+| y_2||^2+\cdots\bigg)^{\frac{1}{2}}\ <\ \infty
\eea
and function $T\rightarrow\ |\lag Tx_1, y_1\rag + \lag Tx_2, y_2\rag+\cdots|$ is a seminorm in $L(H)$. A collection of all these seminorms defines the Hausdorff locally convex topology in $L(H)$ called ultra-weak topology.

\subsection{The Borel Spaces and Measure}
Let $z$ be the Borel space, and a set $B$ of Borel sets of $z$. A subset of $z$ is said to be $\nu$-negligible if it is contained in a set $Y\in B$ such that $\nu(Y)=0$. A subset of $z$ is said to be $\nu$-measurable if it is the form $X\cup N$, where $X\in B$ and $N$ is $\nu$-negligible. A field of complex Hilbert spaces over $z$ is a mapping $\xi\rightarrow H(\xi)$ defined on $z$. $H(\xi)$ is a complex Hilbert space for $\xi\in z$. A measure is said to be standard if there exists a $\nu$-negligible subset $N$ of $z$ such that the Borel space $z\diagdown N$ is standard.
\begin{theorem}
Suppose that $X$ is standard. If
\bea
A=\int^{\bigoplus} A(x)\ d\nu(x),
\eea
then
\bea
A^{\prime}=\int^{\bigoplus} A^{\prime}(x)\ d\nu(x).
\eea
\end{theorem}

\begin{definition}
We say that $H(\xi)$ form a $\nu$-measurable field of complex Hilbert spaces if there is a linear subspace $S$ of $F$ with following properties:
\nn\\
(1) For each $x\in S$, function $\xi\rightarrow || x(\xi)||$ is $\nu$-measurable,
\nn\\
(2) If $y\in F$, the complex valued function $\xi\rightarrow \lag x(\xi), y(\xi)\rag$ is $\nu$-measurable for each $x\in S$, then $y\in S$,
\nn\\
(3)There exists a sequence $(x_1, x_2, \cdots)$ of elements of $S$. For each $\xi\in z$, $x_n(\xi)$ form a total sequence in $H(\xi)$.
\end{definition}

The constant field corresponding to $H_0$ (separable complex Hilbert space) over $z$ is the $\nu$-measurable field defined as: a. $H(\xi)=H_0$ for every $\xi\in z$, b. the $\nu$-measurable vector fields are the $\nu$-measurable mappings of $z$ into $H_0$.

\subsection{Some Useful Theorems of the Von-Neumann Algebra}
We show some useful theorems of the Von-Neumann algebra in the entanglement entropy, and compute the entanglement entropy from these theorems.
\begin{theorem}
Let $A$ be a *-algebra of operators in $H$ with $X_H^A=H$ ($I_H\in A$, where $I_H$ is an identity operator in $H$).). Then $A^{\prime\prime}$ (the Von-Neumann algebra generated by $A$) is the closure of $A$ in the weak, strong, ultra-weak or ultra-strong topologies.
\end{theorem}
\begin{proof}[Proof]
 Let $A_1$ be the closure of $A$ in the weak topology, which is a *-algebra of operators. Therefore, $A^{\prime}$ and $A^{\prime\prime}$ are weakly closed. This shows $A\subseteq A_1\subseteq A^{\prime\prime}\subseteq A_1^{\prime\prime}$. We also have $A_1=A_1^{\prime\prime}$ due to $I_H\in A$. Therefore, we obtain $A_1=A^{\prime\prime}$. Similarly, we can also show that the closure of $A$ is $A^{\prime\prime}$ in the strong, ultra-weak or ultra-strong topologies.
\end{proof}

This theorem shows existence of the Von-Neumann algebra in quantum field theory. In quantum theory, we have closed and unitary as in the Von-Neumann algebra. Therefore, the Von-Neumann algebra is useful in quantum field theory which do not lose completeness.

\begin{theorem}
\label{diag1}
Suppose $A$ and $A^{\prime}$ are decomposable ($A$ and $A^{\prime}$ are the Von-Neumann algebras.), then
\bea
A=\int^{\bigoplus}A(\xi)\ d\nu(\xi), \qquad A^{\prime}=\int^{\bigoplus}A^{\prime}(\xi)\ d\nu(\xi).
\eea
(1) $A(\xi)$ and $A^{\prime}(\xi)$ commute almost everywhere.
\nn\\
(2) If $Z$ is diagonalizable, and $Z$ is the center of $A$ (and $A^{\prime}$), then $A(\xi)$ and $A^{\prime}(\xi)$ generate the Von-Neumann algebra $L(H(\xi))$.
\nn\\
(3) Conversely, if $A(\xi)$ and $A^{\prime}(\xi)$ generate the Von-Neumann algebra $L(H(\xi))$, then $Z$ is the center of $A$ and $A^{\prime}$, and $A(\xi)$ and $A^{\prime}(\xi)$ are factors.
\end{theorem}
\begin{proof}[Proof]
\bea
T_1&=&\int^{\bigoplus} T_1(\xi)\ d\nu(\xi), \qquad T_2=\int^{\bigoplus}T_2(\xi)\ d\nu(\xi), \cdots,
\nn\\
T_1^{\prime}&=&\int^{\bigoplus} T_1^{\prime}(\xi)\ d\nu(\xi), \qquad T_2^{\prime}=\int^{\bigoplus}T_2^{\prime}(\xi)\ d\nu(\xi), \cdots,
\eea
are decomposable operators. As $T_1,\ T_2,\ \cdots$ commute with $T_1^{\prime},\ T_2^{\prime},\ \cdots$ and $T_1^{\prime \dagger},\ T_2^{\prime\dagger},\ \cdots$, and $T_1(\xi),\ T_2(\xi),\ \cdots$ commute with $T_1^{\prime}(\xi),\ T_2^{\prime}(\xi),\ \cdots$ and $T_1^{\prime \dagger}(\xi),\ T_2^{\prime \dagger}(\xi),\ \cdots$, hence, $A(\xi)$ and $A^{\prime}(\xi)$ commute almost everywhere.

$A\cap A^{\prime}=Z$ is the same as saying that $A$ and $A^{\prime}$ generate the Von-Neumann algebra $Z^{\prime}$. Hence, $T_1, \ T_2,\ \cdots$ and $T_1^{\prime},\ T_2^{\prime},\ \cdots$ generate the Von-Neumann algebra $Z^{\prime}$ almost everywhere. This means that $T_1(\xi),\ T_2(\xi),\ \cdots$ and $T_1^{\prime}(\xi),\ T_2^{\prime}(\xi),\ \cdots$ generate $L\big(H(\xi)\big)$.

Finally, if $A(\xi)$ and $A^{\prime}(\xi)$ generate $L(H(\xi))$, we have 
\bea
A(\xi)\cap A^{\prime}(\xi)=\big(A^{\prime}(\xi)\cup A(\xi)\big)^{\prime}=L^{\prime}\big(H(\xi)\big).
\eea
Therefore, $Z$ is the center of $A$ and $A^{\prime}$, and $A(\xi)$ and $A^{\prime}(\xi)$ are also factors.
\end{proof}
This theorem is useful to use the decomposable algebras $A$ and $A^{\prime}$ to generate $L(H)$, then we can use factors to generate the total algebra or we have tensor product decomposition in each subspace. This property is interesting for a generalization of partial trace operator. In local quantum field theory, we use $A$ and $A^{\prime}$ to generate the Von-Neumann algebra, the algebra is decomposable and center is diagonalizable, then our approach will not lose generality.

\begin{proposition}
Let $\xi\rightarrow H(\xi)$ be a $\nu$-measurable field of complex Hilbert space over $z$, where $H=\int^{\bigoplus}H(\xi)\ d\nu(\xi)$. A separable complex Hilbert space is $K_0$, a constant field corresponding to $K_0$ over $z$ is $\xi\rightarrow K(\xi)$, a decomposable Von-Neumann algebra in $H$ is $A=\int^{\bigoplus}A(\xi)\ d\nu(\xi)$, and $B$ is the Von-Neumann algebra in $K_0$. Therefore, we can identify $H\otimes K_0$ as $\int^{\bigoplus}(H(\xi)\otimes K_0)\ d\nu(\xi)$, then we have $A\otimes B=\int^{\bigoplus}(A(\xi)\otimes B)\ d\nu(\xi)$.
\end{proposition}

\begin{corollary}
Let $K_0$ be a separable complex Hilbert space. A constant field corresponding to $K_0$ over $z$ is $\xi\rightarrow K(\xi)$. For each Von-Neumann algebra $B$ in $K_0$, $Z\otimes B$ is isomorphic to $\int^{\bigoplus}B(\xi)\ d\nu(\xi)$, where $Z$ is the diagonalizable operator, with $B(\xi)=B$ for each $\xi\in z$.
\end{corollary}
Finally, we give a lemma to relate a decomposable Hilbert space and decomposable algebra to diagonalizable operators from a transformation. When we define the entanglement entropy with non-trivial centers, we use $A$ and $A^{\prime}$ to generate full algebras. Hence, centers should be diagonalizable. It is useful to compute the entanglement entropy with the non-trivial centers.
\begin{lemma}
\label{diag2}
Let $K_0$ be a separable complex Hilbert space. The Von-Neumann algebra $A_0$ in $K_0$, and a $\nu$-measurable field of complex Hilbert space over $z$ is $\xi\rightarrow H(\xi)$, where
\bea
H=\int^{\bigoplus}H(\xi)\ d\nu(\xi).
\eea 
The algebra of diagonalizable operator is $Z$, a $\nu$-measurable field of the Von-Neumann algebra over $z$ is $\xi\rightarrow A(\xi)\subset L\big(H(\xi)\big)$, and
\bea
A=\int^{\bigoplus}A(\xi)\ d\nu(\xi).
\eea
Suppose that there exists, for each $\xi\in z$, an isomorphism $U(\xi)$ of $H(\xi)$ onto $K_0$ such that $U(\xi)^{-1}A_0U(\xi)=A(\xi)$. Suppose that $\nu$ is standard. Then there exists a transformation from $A$ into $Z\otimes A_0$. 
\end{lemma}
This lemma is to mention that two Hilbert spaces exist a isomorphism, which gives diagonalizable operators. The diagonalizable operator is also a center of the decomposable operator. This implies that the center is isomorphic to a diagonalizable operator. 

It is interesting for finding a transformation to get an algebra with a new structure, which will give equivalent physical results in a total Hilbert space. But the entanglement entropy will depend on mapping so different choices of mapping will give different answers. This can be seen as a new way to extend the entanglement entropy from tensor product decomposition of the Hilbert space to non-tensor product decomposition of the Hilbert space. The entanglement entropy with gauge symmetry is hard to define comes from the partial trace operator in the non-tensor product decomposition of the Hilbert space. When using mapping this new factor, a suitable partial trace operator will have hope to find in the non-tensor product decomposition of the Hilbert space. Using transformation to define the entanglement entropy can be seen as changing observation way. We will find a basis to diagonalize center, and let algebra be a block diagonal form. This can be done generically in local quantum field theory.

\section{Details of the Strong Subadditivity}
We show all lemmas and theorems related to the strong subadditivity \cite{Araki:1970ba, Lieb:1973cp} in this appendix. 
\label{app3}
\begin{lemma}
\label{inequality for log}
\bea
\ln x\le x-1 \qquad x> 0
\eea
with equality only at x=1.
\end{lemma}

\begin{lemma}
\label{Klein inequality}
If $A$ and $B$ are self-adjoint positive trace-class operators on the Hilbert space $H$, then
\bea
\mathrm{Tr}\bigg(A\ln A-A\ln B-A+B\bigg)\ge 0.
\eea
\end{lemma}
\begin{proof}[Proof]
Let $\psi_i$ ($\phi_i$) be a complete orthonormal set of $A$ ($B$) with eigenvalues $a_i$ ($b_i$). The relation between $\psi_i$ and $\phi_i$ is given by
\bea
\phi_i=\sum_j U_{ij}\phi_j,
\eea
where $U_{ij}$ is a unitary mapping. Therefore,
\bea
\lag\psi_i| A\ln A- A\ln B|\psi_i\rag&=&a_i\bigg(\ln a_i-\sum_j| U_{ij}|^2\ln b_j\bigg)\ge 
a_i\bigg\lbrack\ln a_i-\ln\bigg(\sum_j| U_{ij}|^2b_j\bigg)\bigg\rbrack
\nn\\
&\ge&a_i-\sum_j|U_{ij}|^2b_j=\lag\psi_i|A-B|\psi_i\rag,
\eea
where we used
\bea
\ln\bigg(\sum_j|U_{ij}|^2b_j\bigg)\ge\sum_j|U_{ij}|^2\ln b_j
\eea
in the first inequality and the Lemma \ref{inequality for log} in the second inequality. Now we find
\bea
\mbox{Tr}\bigg(A\ln A-A\ln B-A+B\bigg)\ge 0.
\eea
\end{proof}

For convenience, we will give some basic definitions and derivations before we derive next mathematical results. We define that $f(A)$ is convex if 
\bea
f\bigg(\lambda A+(1-\lambda)B\bigg)\le\lambda f(A)+(1-\lambda) f(B) \qquad 0\le\lambda\le1
\eea
and $f(A)$ is concave if
\bea
f\bigg(\lambda A+(1-\lambda)B\bigg)\ge\lambda f(A)+(1-\lambda) f(B) \qquad 0\le\lambda\le1.
\eea
If $f(A)$ is concave, then $-f(A)$ is convex. Now we introduce integral representation of $\ln a$ function, where $a$ is a number, as 
\bea
\int_0^{\infty} dx\ (1+x)^{-1}(a-1)(a+x)^{-1}=\int_0^{\infty}dx\ \bigg(\frac{1}{1+x}-\frac{1}{x+a}\bigg)=\ln a.
\eea
If $A$ is a bounded self-adjoint strictly positive ($\lag x, Ax\rag\ge 0\ \forall x\neq 0$) linear operator, then we have 
\bea
\ln A=\int_0^{\infty}dy\ (1+y)^{-1}(A-1)(A+y1)^{-1},
\eea
which can be shown by choosing a basis to let $A$ be diagonal. The procedure of proof is similar with the integral representation of $\ln a$. Now we use the integral representation to show
\bea
\frac{d}{dx}\ln(\alpha+x\beta)|_{x=0}\equiv T_{\alpha}(\beta)\equiv \int_0^{\infty} dy\ (\alpha+y1)^{-1}\beta(\alpha+y1)^{-1}
\eea
where $\alpha$ is a bounded self-adjoint strictly positive and $\beta$ is a bounded self-adjoint linear operators.
\bea
\ln(\alpha+x\beta)&=&\int_0^{\infty}dy\ (1+y)^{-1}(\alpha+x\beta-1)(\alpha+x\beta+y1)^{-1},
\nn\\
\frac{d}{dx}\ln(\alpha+x\beta)&=&\int_0^{\infty}dy\ (1+y)^{-1}\beta(\alpha+y1+x\beta)^{-1}
\nn\\
&&-\int_0^{\infty}dy\ (1+y)^{-1}(\alpha+x\beta-1)(\alpha+y1+x\beta)^{-1}\beta(\alpha+y1+x\beta)^{-1},
\nn\\
\frac{d}{dx}\ln(\alpha+x\beta)|_{x=0}&=&\int_0^{\infty}dy\ (1+y)^{-1}\beta (\alpha+y1)^{-1}
\nn\\
&&-\int_0^{\infty} dy\ (1+y)^{-1}(\alpha -1)(\alpha+y1)^{-1}\beta(\alpha+y1)^{-1}
\nn\\
&=&\int_0^{\infty}dy\ (\alpha+y1)^{-1}\beta(\alpha+y1)^{-1}.
\eea
Later we will need the second order derivative so we compute it now as 
\bea
\frac{d^2}{dx^2}\ln(\alpha+x\beta)|_{x=0}&=&-2\int_0^{\infty}dy\ (1+y)^{-1}\beta(\alpha+y1)^{-1}\beta(\alpha+y1)^{-1}
\nn\\
&&+2\int_0^{\infty}dy\ (1+y)^{-1}(\alpha-1)(\alpha+y1)^{-1}\beta(\alpha+y1)^{-1}\beta(\alpha+y1)^{-1}
\nn\\
&=&-2\int_0^{\infty}dy\ (\alpha+y1)^{-1}\beta(\alpha+y1)^{-1}\beta(\alpha+y1)^{-1}\equiv-R_{\alpha}(\beta).
\nn\\
\eea
\begin{lemma}
\label{convex inequalities}
Let $F$ be a convex function, and its derivative is
\bea
\lim_{x\rightarrow 0}x^{-1}\bigg(F(A+xB)-F(A)\bigg)\equiv G(A, B).
\eea
Assume that $F$ is homogeneous or order 1 ($F(\lambda A)=\lambda F(A)$ for $\lambda >0$.). Then we have $G(A, B)\le F(B)$.
\end{lemma}
\begin{proof}[Proof]
For all $x>0$,
\bea
F(A+xB)&=&F\bigg\lbrack (1+x)\bigg((1+x)^{-1}A+x(1+x)^{-1}B\bigg)\bigg\rbrack
\nn\\
&=&(1+x)F\bigg((1+x)^{-1}A+x(1+x)^{-1}B\bigg)
\nn\\
&\le&(1+x)\bigg((1+x)^{-1}F(A)+x(1+x)^{-1}F(B)\bigg)=F(A)+xF(B).
\nn\\
\eea
Therefore, we obtain
\bea
G(A, B)\le F(B).
\eea
\end{proof}
A convex function is $Q(A, K)=\mbox{Tr}\bigg(K T_A(K)\bigg)$ (This is homogeneous of order 1.), and its derivative is given by
\bea
\frac{d}{dx}T_{A+xB}(L)&=&\frac{d}{dx}\int_0^{\infty}dy\ (A+y1+xB)^{-1}K(A+y1+xB)^{-1}\bigg|_{x=0}
\nn\\
&=&-\int_0^{\infty}dy\ (A+y1)^{-1}B(A+y1)^{-1}K(A+y1)^{-1}
\nn\\
&&-\int_0^{\infty}dy\ (A+y1)^{-1}K(A+y1)^{-1}B(A+y1)^{-1},
\nn\\
\frac{d}{dx}Q(A+xB, K)|_{x=0}&=&-\mbox{Tr}\bigg(BR_A(K)\bigg). 
\eea
By using the Lemma \ref{convex inequalities}, $Q(A, K)$ gives
\bea
\label{inequalities for Q}
-\mbox{Tr}\bigg(BR_A(K)\bigg)+2\mbox{Tr}\bigg(MT_A(K)\bigg)
\le\mbox{Tr}\bigg(MT_B(M)\bigg),
\eea
where $A$ and $B$ are bounded self-adjoint strictly positive linear operators and $K$ is a bounded self-adjoint linear operator.
\begin{theorem}
Let $L$ be a bounded self-adjoint positive ($\lag x, Ax\rag\ge 0$ $\forall x$) and $A$ be a bounded self-adjoint strictly positive linear operators. Then
\bea
A\mapsto F_L(A)\equiv \mathrm{Tr}\bigg(\exp(L+\ln A)\bigg)
\eea
is concave for all $L$.
\end{theorem}
\begin{proof}[Proof]
Choose $K$ to be a bounded self-adjoint linear operator, and define $f(x)\equiv\mbox{Tr}\bigg\lbrack\exp\bigg(L+\ln\big(A+xK\big)\bigg)\bigg\rbrack$. This theorem is equivalent to showing $\frac{d^2 f}{d^2 x}\le 0$ when $x=0$ for all $A$, $L$ and $K$.  Then the computation is given by
\bea
\frac{d}{dx}e^{F+xG}\bigg|_{x=0}&=&\int_0^1dy\ e^{yF}Ge^{(1-y)F}=T^{-1}_{\exp(F)}(G),
\nn\\
\frac{d^2f}{dx^2}\mid_{x=0}&=&-\mbox{Tr}\bigg(BR_A(K)\bigg)+\mbox{Tr}\bigg(\int_0^1dy\ T_A(K)B^yT_A(K)B^{1-y}\bigg)
\nn\\
&=&-\mbox{Tr}\bigg( BR_A(K)\bigg)+\mbox{Tr}\bigg\lbrack T_A(K)T_B^{-1}\bigg(T_A(K)\bigg)\bigg\rbrack,
\eea
where $B=\exp(L+\ln A)$. We use \eqref{inequalities for Q} by choosing $M=T_B^{-1}\bigg(T_A(K)\bigg)$ to show $\frac{d^2f}{dx^2}\le0$. Therefore, we obtain that $F_L(A)$ is concave for all $L$.
\end{proof}
\begin{theorem}
\label{Golden-Thompson inequality}
Let $A$, $B$ and $C$ be bounded self-adjoint operators, then
\bea
\mathrm{Tr}\bigg(e^CT_{\exp(-A)}\big(e^B\big)\bigg)\ge\mathrm{Tr}\bigg(e^{A+B+C}\bigg).
\eea
\end{theorem}
\begin{proof}[Proof]
We define $\alpha\equiv e^{-A}$, $\beta\equiv e^B$ and $L\equiv A+C$. Then we use the Lemma \ref{convex inequalities} to deduce
\bea
\mbox{Tr}\bigg( e^{A+B+C}\bigg)=\mbox{Tr}\bigg(e^{L+\ln\beta}\bigg)\le\frac{d}{dx}\mbox{Tr}\bigg(e^{L+\ln(\alpha+x\beta)}\bigg)\bigg|_{x=0}=\mbox{Tr}\bigg(e^CT_{\alpha}\big(\beta\big)\bigg).
\eea
\end{proof}

\section{Quantum Entropy in Free Theory}
\label{app4}
We explicitly compute quantum entropy in free theory in this appendix, and start from the density matrix
\bea
\rho_B=Ke^{-\sum_{l_B}\epsilon_{l_B}a_{l_B}^{\dagger}a_{l_B}},
\eea
where $K$ is a normalization constant, and $a_{i_B}$ and $a_{i_B}^{\dagger}$ is defined as
\bea
\phi_{i_B}=\alpha_{i_Bj_B}^*a_{j_B}^{\dagger}+\alpha_{i_Bj_B}a_{j_B}, \qquad \pi_{i_B}=-i\beta_{i_Bj_B}^*a_{j_B}^{\dagger}+i\beta_{i_Bj_B}a_{j_B}
\eea 
with
\bea
\lbrack a_{i_B}, a_{j_B}^{\dagger}\rbrack=\delta_{i_Bj_B}, \qquad \alpha^*\beta^T+\alpha\beta^{\dagger}=-1.
\eea
The normalization $K$ can be determined as
\bea
1=K\mbox{Tr}\bigg(e^{-\sum_{l_B}\epsilon_{l_B}a_{l_B}^{\dagger}a_{l_B}}\bigg)=
K\Pi_{k_B}\bigg(1+e^{-\epsilon_{k_B}}
+e^{-2\epsilon_{k_B}}+\cdots\bigg)=K\Pi_{k_B}\frac{1}{1-e^{-\epsilon_{k_B}}}.
\nn\\
\eea
Hence, $K=\Pi_{l_B}\bigg(1-e^{-\epsilon_{l_B}}\bigg)$. Then we show expectation value of $a_{k_B}^{\dagger}a_{k_B}$ as
\bea
n_{k_Bk_B}\equiv\lag a_{k_B}^{\dagger}a_{k_B}\rag=\frac{\partial}{\partial\epsilon_{k_B}}\ln(1-e^{-\epsilon_{k_B}})=\frac{1}{e^{\epsilon_{k_B}}-1}.
\eea
Other expectation values in $a_{k_B}$ and $a_{k_B}^{\dagger}$ are
\bea
\lag a_{k_B}^{\dagger}a_{l_B}\rag=n_{k_Bk_B}\delta_{k_Bl_B}, \qquad \lag a_{k_B}^{\dagger}a_{l_B}^{\dagger}\rag=0, \qquad \lag a_{k_B}a_{l_B}\rag=0.
\eea
From
\bea
\mbox{Tr}(\rho_B q_{i_B}p_{j_B})=\frac{i}{2}\delta_{i_Bj_B}, \qquad \mbox{Tr}(\rho_Bq_{i_B}q_{j_B})=X_{i_Bj_B}, \qquad \mbox{Tr}(\rho_Bp_{i_B}p_{j_B})=P_{i_Bj_B},
\nn\\
\eea
the first equality gives
\bea
\label{1}
\alpha^*n\beta^T-\alpha(n+1)\beta^{\dagger}=\frac{1}{2}1,
\eea
the second equality gives
\bea
\label{2}
\alpha^*n\alpha^T+\alpha(n+1)\alpha^{\dagger}=X^B,
\eea
and the third equality gives
\bea
\label{3}
\beta^*n\beta^T+\beta(n+1)\beta^{\dagger}=P^B,
\eea
where $\alpha\equiv\alpha_{i_Bj_B}$, $n\equiv n_{i_Bj_B}$, $\beta\equiv\beta_{i_Bj_B}$, $X^B\equiv X_{i_Bj_B}$ and $P^B\equiv P_{i_Bj_B}$. Solving these equations \eqref{1}-\eqref{3}, we can redefine $a_i$ to let $\alpha$ and $\beta$ be real. Hence, \eqref{1} gives 
\bea
\label{4}
\alpha=-\frac{1}{2}(\beta^T)^{-1}.
\eea 
Then \eqref{2}-\eqref{4} gives
\bea
\alpha\frac{1}{4}(2n+1)^2\alpha^{-1}=X^BP^B.
\eea
Now we can obtain $\frac{1}{4}(2n_{k_Bk_B}+1)^2=\nu_{k_B}^2$, where $\nu_{k_B}$ are eigenvalues of $\sqrt{X^BP^B}$. Hence, $\tanh(\frac{\epsilon_{l_B}}{2})=\frac{1}{2\nu_{l_B}}$. The quantum entropy is given by
\bea
S_Q(V)&=&-\mbox{Tr}(\rho_B\ln\rho_B)=-\sum_{l_B}\ln (1-\epsilon^{-\epsilon_{l_B}})+\sum_{l_B}\epsilon_{l_B}\lag a_{l_B}^{\dagger}a_{l_B}\rag
\nn\\
&=&-\sum_{l_B}\bigg(\ln(1-e^{-\epsilon_{l_B}})+\frac{\epsilon_{l_B} e^{-\epsilon_{l_B}}}{1-e^{-\epsilon_{l_B}}}\bigg)
\nn\\
&=&=\sum_{l_B}\bigg((\nu_{l_B}+\frac{1}{2})\ln(\nu_{l_B}+\frac{1}{2})-(\nu_{l_B}-\frac{1}{2})\ln(\nu_{l_B}-\frac{1}{2})\bigg)
\nn\\
&=&\mbox{Tr}\bigg((C+\frac{1}{2})\ln(C+\frac{1}{2})-(C-\frac{1}{2})\ln(C-\frac{1}{2})\bigg),
\nn\\
\eea
where we used
\bea
-\ln(1-e^{-\epsilon_{l_B}})=\ln(\nu_{l_B}+\frac{1}{2}), \qquad \frac{\epsilon_{l_B} e^{-\epsilon_{l_B}}}{1-e^{-\epsilon_{l_B}}}=(\nu_{l_B}-\frac{1}{2})\ln\bigg(\frac{\nu_{l_B}+\frac{1}{2}}{\nu_{l_B}-\frac{1}{2}}\bigg),
\eea
and $C\equiv(X^BP^B)^{1/2}$.

\section{Details of the Entanglement Entropy in the Einstein Gravity Theory}
\label{app5}
The entanglement entropy is
\bea
S_{\mbox{EE}}=-\partial_n\bigg(\ln Z_n-nZ_1\bigg)\bigg|_{n=1},
\eea
where the partition function in the Einstein gravity theory will behave like the delta function near entangling surface. The classical partition function in the Einstein gravity theory is defined as $Z=\exp(-S)$, where $S$ is the action. 

When we take on-shell solutions into the Einstein gravity theory, the on-shell bulk action does not have derivative terms. Therefore, we should not have any conical singularity on the bulk because our metric is periodic with respect to conical angle. Hence, the entanglement entropy in the Einstein gravity theory only comes from the entangling surface. When we consider the higher derivative gravity theory, the bulk on-shell action possibly have derivative terms. Hence, we also need to consider bulk entanglement entropy in the higher derivative theory. 

 Now we use conical singularity to find the entanglement entropy on the entangling surface. We set the entangling surface at $\rho=0$ and conical angle $2\pi n$, and start from an off-shell metric field near $\rho=0$ \cite{Ryu:2006bv, Lewkowycz:2013nqa, Fursaev:1995ef}
\bea
ds^2=\bigg(1+\frac{(n^2-1)a^2}{\rho^2+a^2}\bigg)d\rho^2+n^2\rho^2d\phi^2,
\eea
where $g_{\rho\rho}\equiv\bigg(1+\frac{(n^2-1)a^2}{\rho^2+a^2}\bigg)$ and $g_{\phi\phi}\equiv n^2\rho^2$. We also have limit
\bea
g_{\rho\rho}|_{\rho=0}=n^2, \qquad g_{\rho\rho}|_{a\rightarrow 0}=1.
\eea
For $n=1$ and $a\rightarrow 0$, we have asymptotic flat metric. Now we analyze singularity. The computation is given by
\bea
\partial_{\rho}g_{\rho\rho}=-\frac{2(n^2-1)a^2\rho}{(\rho^2+a^2)^2}, \qquad R=\frac{1}{g_{\rho\rho}}R_{\rho\rho}+\frac{1}{g_{\phi\phi}}R_{\phi\phi}, \qquad \sqrt{\det{g_{\mu\nu}}}=\sqrt{g_{\rho\rho}g_{\phi\phi}},
\eea
where
\bea
\Gamma^{\mu}_{\nu\delta}=\frac{1}{2}g^{\mu\lambda}\bigg(\partial_{\delta}g_{\lambda\nu}+\partial_{\nu}g_{\lambda\delta}
-\partial_{\lambda}g_{\nu\delta}\bigg), 
\qquad
R_{\mu\nu}=\partial_{\delta}\Gamma^{\delta}_{\nu\mu}-\partial_{\nu}\Gamma^{\delta}_{\delta\mu}
+\Gamma^{\delta}_{\delta\lambda}\Gamma^{\lambda}_{\nu\mu}
-\Gamma^{\delta}_{\nu\lambda}\Gamma^{\lambda}_{\delta\mu}.
\nn\\
\eea
\bea
\Gamma^{\rho}_{\rho\rho}=-\frac{(n^2-1)a^2\rho}{(\rho^2+a^2)(\rho^2n^2+a^2)}, \qquad
\Gamma^{\rho}_{\phi\phi}=-\bigg(\frac{\rho^2+a^2}{\rho^2+n^2a^2}\bigg)n^2\rho, \qquad
\Gamma^{\phi}_{\rho\phi}=\frac{1}{\rho}, 
\nn\\
\eea
\bea
\Gamma^{\rho}_{\rho\phi}=\Gamma^{\phi}_{\rho\rho}=\Gamma^{\phi}_{\phi\phi}=0.
\eea
Hence,
\bea
R_{\rho\rho}=-\frac{(n^2-1)a^2}{(\rho^2+a^2)(\rho^2+n^2a^2)}, \qquad R_{\phi\phi}=\frac{\rho^2a^2(1-n^2)}{(\rho^2+a^2n^2)^2}n^2, \qquad R=\frac{2a^2(1-n^2)}{(\rho^2+a^2n^2)^2},
\nn\\
\eea
\bea
\sqrt{\det g_{\mu\nu}}=\sqrt{\frac{\rho^2+a^2n^2}{\rho^2+a^2}}n\rho.
\eea
The contribution of the action in $D$-dimensions from the entangling surface as 
\bea
-\frac{1}{16\pi G}\int d^Dx\sqrt{\det{g_{\mu\nu}}}R
=-\frac{1}{16\pi G}\int d^{D-2}x\lim_{a\rightarrow 0}4\pi n\int_{0}^{\epsilon}\frac{a^2(1-n^2)}{(\rho^2+a^2n^2)^{\frac{3}{2}}(\rho^2+a^2)^{\frac{1}{2}}}\rho d\rho,
\nn\\
\eea
where $\epsilon$ is a small parameter (But we let $\epsilon/ a\gg 1$.) and $G$ is the Newton constant. We use change variable 
\bea
h^3\equiv\bigg(\frac{\rho^2+a^2n^2}{\rho^2+a^2}\bigg)^{\frac{3}{2}}, \qquad 
\frac{dh^2}{d\rho}=-\frac{2(n^2-1)a^2\rho}{(\rho^2+a^2)^2}
\eea
to compute the integral
\bea
&&\lim_{a\rightarrow 0}4\pi n\int_{0}^{\epsilon}\frac{a^2(1-n^2)}{(\rho^2+a^2n^2)^{\frac{3}{2}}(\rho^2+a^2)^{\frac{1}{2}}}\rho d\rho=4\pi n\int_0^{\epsilon}\frac{\frac{dh^2}{d\rho}}{2h^3}d\rho=-4\pi n h^{-1}\bigg|_0^{\epsilon}
\nn\\
&=&-\frac{4\pi n}{\sqrt{\rho^2+a^2n^2}}\sqrt{\rho^2+a^2}\bigg|_0^{\epsilon}=4\pi(1-n).
\eea
Therefore, the entanglement entropy is $A_{q-2}/(4G)$ \cite{Ryu:2006bv, Lewkowycz:2013nqa, Fursaev:1995ef}, where $A_{q-2}$ is codimension two surface.  This result does not change even if we embed two dimensional cone in higher dimensions because the conical singularity only appears in the two dimensional cone (The analysis method is similar with what we did.). The above computation is to use a two dimensional off-shell cone with a parameter $a$ to compute the entanglement entropy in the Einstein gravity theory. We can also start from a two dimensional on-shell cone to compute the entanglement entropy. Then we need to put a boundary term in the on-shell cone method. The entanglement entropy only comes from this boundary term and the entanglement entropy on the bulk will vanish. The answer is also consistent with the off-shell cone method. The off-shell cone is not a rigorous method naively. When you use an off-shell cone method to compute the entanglement entropy, the derivative of the metric field is ill-defined when you take limit. Hence, the off-shell cone method is just a way to do regularization for the delta function. In other words, the off-shell cone method is to find the boundary effect on the bulk action from the regularization without putting a boundary term. Therefore, we can use the on-shell cone to compute the entanglement entropy with a boundary term, or we use the off-shell cone to do regularization without putting a boundary term. 

\section{$AdS_5$ Metric}
\label{app6}
The $AdS_5$ metric is
\bea
ds^2=l_s^2\bigg(\frac{1}{\sqrt{\lambda}}\frac{-dt^2+\sum_{i=1}^3 dx_i^2}{Z^2}+\sqrt{\lambda}\frac{dZ^2}{Z^2}\bigg),
\eea
where $\lambda\equiv 4\pi g_sN$, $Z\equiv l_s^2/r$, $l_s$ is string length, $g_s$ string coupling constant, $r$ is distance that parallel D3-branes separated and $N$ is rank of the gauge group \cite{Maldacena:1997re}. We can rewrite the metric as
\bea
ds^2=\frac{l_s^2\sqrt{\lambda}}{Z^{\prime 2}}\bigg(-dt^2+\sum_{i=1}^3dx^2+dZ^{\prime 2}\bigg),
\eea
where $Z^{\backprime}=\sqrt{\lambda}Z$. Then we use $Z^{\backprime\backprime}=1/Z^{\backprime}$ and $Z^{\backprime\backprime\backprime}=Z^{\backprime\backprime}l_s^2\sqrt{\lambda}$ to get
\bea
ds^2=l_s^2\sqrt{\lambda}\bigg(\frac{dZ^{\backprime\backprime 2}}{Z^{\backprime\backprime 2}}+Z^{\backprime\backprime 2}\big(-dt^2+\sum_{i=1}^3dx_i^2\big)\bigg)
=l_s^2\sqrt{\lambda}\frac{dZ^{\backprime\backprime\backprime 2}}{Z^{\backprime\backprime\backprime 2}}+\frac{Z^{\backprime\backprime\backprime 2}}{l_s^2\sqrt{\lambda}}(-dt^2+\sum_{i=1}^3dx_i^2).
\nn\\
\eea
Now we use
\bea
X_0=tZ^{\backprime\backprime\backprime}l_s\lambda^{\frac{1}{4}}, \qquad X_{i^{\prime}}=x_{i^{\prime}} Z^{\prime\prime\prime} l_s\lambda^{\frac{1}{4}}, 
\eea
\bea
X_4+X_5=Z^{\backprime\backprime\backprime}, \qquad
 X_5-X_4=\frac{(-t^2+\sum_{i=1}^3X_i^2)Z^{\backprime\backprime\backprime}}{l_s^2\sqrt{\lambda}}+\frac{l_s^2\sqrt{\lambda}}{Z^{\backprime\backprime\backprime}},
\eea
where $i^{\prime}=1, 2, 3$. Hence, the metric is
\bea
ds^2=-dX_0^2-dX_5^2+dX_1^2+dX_2^2+dX_3^2+dX_4^2
\eea
with
\bea
X_0^2+X_5^2-X_1^2-X_2^2-X_3^2-X_4^2=\sqrt{\lambda}l_s^2.
\eea 
Then we use
\bea
X_0=l_s\lambda^{\frac{1}{4}}\cosh\rho \cos\tau, \qquad X_5=l_s\lambda^{\frac{1}{4}}\cosh\rho \sin\tau, \qquad X_{i^{\prime\prime}}=l_s\lambda^{\frac{1}{4}}\sinh\rho\hat{x}_{i^{\prime\prime}},
\eea
where $\sum_{i=1}^4\hat{x}_i^2=1$, and indices $i^{\prime\prime}=1, 2, 3, 4$, to obtain
\bea
ds^2=l_s^2\lambda^{\frac{1}{2}}\bigg(-\cosh^2\rho\ d\tau^2+d\rho^2
+\sinh^2\rho\ d\Omega^2_3\bigg),
\eea
where
\bea
d\Omega_3^2=d\theta^2+\sin^2\theta(d\psi^2+\sin^2\psi d\phi^2).
\eea
Now we use 
\bea
\tilde{r}=l_s^2\sqrt{\lambda}\sinh\rho, \qquad \tilde{t}=l_s^2\sqrt{\lambda}\tau
\eea
to get
\bea
ds^2=-\bigg(1+\frac{\tilde{r}^2}{l_s^4\lambda}\bigg)d\tilde{t}^2+\frac{1}{1+\frac{\tilde{r}^2}{l_s^4\lambda}}d\tilde{r}^2+\tilde{r}^2d\Omega_3^2.
\eea
When we take $N\rightarrow \infty$ and $g_sN \gg 1$, the metric will approach to flat metric as \cite{Maldacena:1997re}
\bea
ds^2=-d\tilde{t}^2+d\tilde{r}^2+\tilde{r}^2d\Omega_3^2.
\eea

\section{Hamitonian Formulation in the Lattice $SU(N)$ Yang-Mills Gauge Theory in the Fundamental Representation}
\label{app7}
We start from the Hamitonian formulation of the $SU(N)$ Yang-Mills gauge theory in the fundamental representation on continuous space. Then we give a Hamitonian formulation on lattice. When we take continuum limit, the lattice model will become the $SU(N)$ Yang-Mills gauge theory in the fundamental representation on continuous space. The action for the $SU(N)$ Yang-Mills gauge theory in the fundamental representation is given by
\bea
S_{\mbox{YMF}}=-\int d^Dx\ \frac{1}{4g^2}F_{\mu\nu}^aF^{\mu\nu,a},
\eea
where
\bea
F_{\mu\nu}^a\equiv\partial_{\mu}A_{\nu}^a-\partial_{\nu}A_{\mu}^a+f^{abc}A_{\mu}^bA_{\nu}^c,
\eea
$A_{\mu}^a$ is one-form gauge potential, and $F_{\mu\nu}^a$ is field strength associated with $A_{\mu}^a$. We denote total spacetime indices by the Greek letters and the Lie algebra indices as $a$-$h$. 
An equation of motion is
\bea
D_{\mu}F^{\mu\nu,a}\equiv\partial_{\mu}F^{\mu\nu,a}+f^{abc}A_{\mu}^bF^{\mu\nu,c}=0,
\eea
where $D_{\mu}$ is covariant derivative. The Lagrangian is given by
\bea
L_{\mbox{YMF}}\equiv\int d^Dx\ \bigg(\frac{1}{2g^2}F_{0i}^aF_{0i}^a-\frac{1}{4g^2}F_{ij}^aF_{ij}^a\bigg).
\eea
We used a Minkowski-space metric with signature $(-1, 1, 1, \cdots).$
Canonical momenta are given by
\bea
\pi_0^a\equiv\frac{\delta L}{\delta \dot{A_0^a}}=0, \qquad \pi_i^a\equiv\frac{\delta L}{\delta \dot{A_i^a}}=\frac{1}{g^2}F_{0i}^a.
\eea
Due to $\pi_0^a=0$, we cannot obtain $\lbrack A_0^a, \pi_0^b\rbrack=\delta^{ab}\delta(\vec{x}-\vec{y})$. With suitable canonical commutation relations, we need to eliminate $A_0^a$. This is the temporal gauge $A_0^a=0$. The Hamitonian in the temporal gauge is given by
\bea
H_{\mbox{YMF}}=\int d^{D-1}x\ \bigg(\pi_{i}^a\partial_0A_{i}^a-L_{\mbox{YMF}}\bigg)=\int\ d^{D-1}x\ \bigg(\frac{g^2}{2}\pi_i^a\pi_i^a+\frac{1}{4g^2}F_{ij}^aF_{ij}^a\bigg).
\nn\\
\eea
From the equation of motion, we have one constraint
\bea
T^a\equiv\partial_i\pi_i^a+f^{abc}A_i^b\pi_i^c=0.
\eea
This is the constraint equation for every spacetime point. Imposing this at one time, this constraint is compatible with the Hamilton's equation. This question can be answered in the quantized case. Assuming the canonical commutation relations as
\bea
\lbrack A_i^a(\vec{x}), \pi_j^b(\vec{y})\rbrack=\delta_{ij}\delta^{ab}\delta(\vec{x}-\vec{y}), \qquad \lbrack\pi_i^a(\vec{x}), \pi_j^b(\vec{y})\rbrack=0, \qquad \lbrack A_i^a(\vec{x}), A_j^b(\vec{y})\rbrack=0.
\nn\\
\eea 
The constraint equations $T^a$ generate time-independent gauge transformation, and $\Omega^{\dagger}\hat{A}_i^a\Omega=$ infinitesimally gauge-transformed $A_i^a$, where $\Omega=1+i\int d^{D-1}x\ \omega^a(\vec{x})T^a(\vec{x})\cdots$. The Hamitonian is gauge invariant so $\lbrack T^a, H\rbrack=0$. The constraint is compatible with the Heisenberg equation of motion. The Hilbert-space realization of the canonical commutation is given by a coordinate representation as
\bea
\lag G|A_i^a(\vec{x})|\Psi\rag=A_i^a(\vec{x})\lag G|\Psi\rag, \qquad \lag G|\pi_i^a(\vec{x})|\Psi\rag=-i\frac{\delta}{\delta A_i^a(\vec{x})}\lag G|\Psi\rag
\eea 
with wavefunctional $\Psi(G)\equiv\lag G|\Psi\rag$. There are no negative norm states, and physical states have to be gauge invariant because we have
\bea
T^a(\vec{x})|\Psi\rag_{\mbox{phys}}=0.
\eea
to define physical states.
Let us write the lattice Hamitonian.
\begin{equation}
H_{\mbox{LYMF}} = \frac{g^2}{2}\sum_l E_l^a E_l^a - \frac{1}{g^2}\sum_\sq\left(\Tr U_\sq + \Tr U_\sq^\hc\right) \,,
\end{equation}
where $E_l^a$ is the electric field operator on spatial link $l$ with component indices $a$ and $U_\sq$ is the plaquette operator on spatial plaquette $\sq$. We set lattice spacing be 1 for simplicity. Now we show that this model have a correct continuum limit. The first term of the lattice Hamitonian gives 
\bea
-\int d^{D-1}x\ \frac{g^2}{2}\frac{\delta}{\delta A_i^{a}}\frac{\delta}{\delta A_i^{a}}
\eea
in continuum limit. Let use check the second term of the lattice Hamitonian.
\bea
U_{ij}U_{jk}=\exp(iA_{m})\exp(iA_{n})=\exp\bigg(iA_{m}+iA_{n}+\frac{i}{2}\lbrack A_{m}, A_{n}\rbrack+\cdots\bigg),
\eea
\bea
U_{kl}U_{li}&=&\exp(-iA^{\prime}_{m})\exp(-iA^{\prime}_{n})=\exp\bigg(-iA^{\prime}_{m}-iA^{\prime}_{n}+\frac{i}{2}\lbrack A_{m}, A_{n}\rbrack+\cdots\bigg)
\nn\\
&=&\exp\bigg(-iA_{m}-iA_{n}-i\partial_{n}A_{m}+i\partial_{m}A_{n}+\frac{i}{2}\lbrack A_{m}, A_{n}\rbrack+\cdots\bigg).
\eea
Therefore, we obtain
\bea
U_{ij}U_{jk}U_{kl}U_{li}=\exp\bigg(iF_{mn}+\cdots\bigg)
\eea
We denote $F_{mn}\equiv F_{mn}^at^a$ and $A_m\equiv A_m^at^a$, where $t^a$ is a generator of the $SU(N)$ group in the fundamental representation. The generator satisfies
\bea
\lbrack t^a, t^b\rbrack=f^{abc}t^c, \qquad \mbox{Tr}(t_a t_b)=\frac{1}{2}\delta_{ab}.
\eea
 Hence, the second term of the lattice Hamitonian gives
\bea
\int d^{D-1}x\ \frac{1}{4g^2}F_{ij}^aF_{ij}^a.
\eea
This show that the lattice Hamitonian has a correct continuum limit. Later, we will use this lattice Hamitonian to study the entanglement entropy. 

\section{Review of the Extended Lattice Model}
\label{app8}
The entanglement entropy of gauge theory is hard to define due to local symmetry. On lattice, we will suffer more obstacles from preserving gauge symmetry with finite lattice spacing. We have a plaquette field which has non-locality on lattice. The plaquette field does not live on the vertices as in lattice scalar field theory. The entanglement entropy in lattice gauge theory will give us these difficulties.

As a remedy, the extended lattice model~\cite{Buividovich:2008gq, Donnelly:2011hn} was proposed, where the original Hilbert space $H$ is embedded into a larger one $H^{\prime}$ that admits a decomposition. This is accomplished by inserting a new vertex on the boundary $\pd A$ at the intersection with a link whenever this crosses $\pd A$ and splitting this link into two new links. The new Hilbert space consists of functions on all the links. The extended lattice model is invariant under gauge transformation acting on all the original vertices, but not on the new vertices. Gauge transformation now acts independently on each side of $\pd A$. Note that $A$ (and similarly $\bar{A}$) now includes all links in its interior, as well as the new links from the boundary. A norm-preserving embedding is given by
\begin{equation}
\label{embedding}
\psi'(U_{l_1},\cdots,U_{l_{\pd A}},U_{l_{\pd\bar{A}}},\cdots,U_{l_k}) \equiv 
\psi(U_{l_1},\cdots,U_{l_{\pd A}} \cdot U_{l_{\pd\bar{A}}},\cdots,U_{l_k}) \,,
\end{equation}
where an embedded state $|\psi'\rag\in H^{\prime}$ is defined by a gauge-invariant state $|\psi\rag\in H$ from the (matrix) dot product of the link variables $U_l$ from all boundary-crossing links $l_\pd \equiv l_{\pd A} \cup l_{\pd\bar{A}}$. Note that the embedded states are invariant even for gauge transformation at the extra or new boundary vertices.

Given a decomposition of $H^{\prime}$, and the embedding, the reduced density matrix and the entanglement entropy are obtained in the standard way:
\begin{equation}
\rho_A = \Tr_{\bar{A}}|\psi'\rag\lag\psi'| \,, \qquad S = -\mbox{Tr}\bigg(\rho_A\log\rho_A\bigg) \,.
\end{equation}
By decomposing the reduced density matrix into irreducible representations of the boundary gauge group, and using properties of the Von-Neumann entropy, the entanglement entropy for a generic state in the lattice gauge theory can be written as a sum of the classical Shannon term, a weighted average involving dimensions of boundary representations, and a term including effects of non-local correlations~\cite{Donnelly:2011hn}. Below we give the formal result and outline their derivation as given in~\cite{Donnelly:2011hn}.

In \cite{Donnelly:2011hn}, a given gauge-invariant state $|\psi\rag\in H$ is expressed in terms of spin network states, which forms an orthonormal basis of $H$:
\begin{equation}
|\psi\rag = \sum_S\psi(S)|S\rag \,,
\end{equation}
where spin network state $|S\rag$ related to a spin network $S$ is a functional obtained by taking  representations $r_l$ of group element on each link $l$, multiplying by $\sqrt{\mbox{dim}(r)}$ and contracting with the interwiner $i_n$ as
\bea
\lag S| U\rag=\bigg(\bigotimes_{l\in L}\sqrt{\mbox{dim}(r_l)}r_l(u_l)\bigg)\cdot\bigg(\bigotimes_{n\in V}i_n\bigg).
\eea 
The interwiner is chosen to be orthonormal in inner product as
\bea
\lag i_1, i_2\rag\equiv\mbox{Tr}(i_1i_2^{\dagger}).
\eea
The resulting spin network states form an orthonormal basis of $H$.
A spin network consists of an assignment of irreducible representations $R = \{r_l: l \in L\}$ to each link, and intertwiners $I = \{i_v: v \in V\}$ to each vertex. Each intertwiner $i_v$ is a gauge invariant map between the representation spaces of all the links that end on the vertex $v$ and all the links that emanate from $v$. A field content is the group element $u_l$ and its gauge transformation is
\bea
u_l\rightarrow g_{t(l)}\cdot u_l\cdot g_{s(l)}^{-1},
\eea
where $g_n$ is a group element, and $s(l)$ and $t(l)$ are nodes at source and target of the link $l$. The map $i_n$ is given by
\bea
i_n:\bigg(\bigotimes_{l:t(l)=n} r_l\bigg)\rightarrow\bigg(\bigotimes_{l:s(l)=n} r_l\bigg).
\eea
A spin network $S$ is then specified by all its representations and intertwiners as
\begin{equation}
S = (R_A,\,R_{\bar{A}},\,R_\pd,\,I_A,\,I_{\bar{A}}) \,.
\end{equation}

Under the embedding $\pi^*:H\ra H^{\prime}_A\otimes H^{\prime}_{\bar{A}}$, the spin network state $|S\rag$ is mapped to~\cite{Donnelly:2008vx}
\begin{equation}\label{eq:Sdecomp}
\pi^*|S\rag = \prod_{l \in L_{\pd A}}\frac{1}{\sqrt{\mathrm{dim}(r_l)}}\sum_{m_l}|S_A\rag\otimes|S_{\bar{A}}\rag \,, 
\end{equation}
where 
\begin{equation}
S_A = (R_A,\,R_\pd,\,I_A,\,M) \,, \qquad S_{\bar{A}} = (R_{\bar{A}},\,R_\pd,\,I_{\bar{A}},\,M^*) \,
\end{equation}
are open spin networks whose states span $H^{\prime}_A$ and $H^{\prime}_{\bar{A}}$ respectively, $M = \{m_l: l \in L_\pd\}$ is a set of vectors in the boundary representation spaces such that $m_l \in r_l$ or its dual $\bar{r}_l$ depending on whether the link $l$ points inward or outward at the boundary, and $M^*$ is the set of vectors dual to those in $M$. Finally, $L_\pd$ is the set of all boundary-crossing links, and $L_{\pd A}$ is a set of all new links created on the side of $A$ when the boundary-crossing links are split. In \eqref{eq:Sdecomp}, the sum over $m_l$ ranges over an orthonormal basis of $r_l$.

Using the decomposition~\eqref{eq:Sdecomp}, the reduced density matrix in region $A$ is given by
\begin{equation}\label{eq:reddm}
\rho_A = \sum_{J}\frac{\psi(S)\psi(S')^*}{\underset{l \in L_{\pd A}}{\prod}\mathrm{dim}(r_l)}|S_A\rag\lag S'_A| \,,
\end{equation}
where
\begin{equation}
S^{\prime} = (R^{\prime}_A,\,R_{\bar{A}},\,R_\pd,\,I^{\prime}_A,\,I_{\bar{A}}) \,, \qquad S^{\prime}_A = (R^{\prime}_A,\,R_\pd,\,I^{\prime}_A,\,M) \,,
\end{equation}
and $J$ is a collective index over $R_{A}$, $R_{\bar{A}}$, $R_{\pd}$, $I_{A}$, $I_{\bar{A}}$, $R^{\prime}_A$, $I^{\prime}_A$, $M$. The sum over intertwiners are over respective orthonormal bases compatible with the representation assignments. Note that $R_\pd$ is the same for both $S_A$ and $S^{\prime}_A$. Thus, $\rho_A$ does not have off-diagonal terms that mix different boundary representations.

To see how the reduced density matrix $\rho_A$ decomposes into representations. Note that $H^{\prime}_A$ has a decomposition
\begin{equation}
H^{\prime}_A = \bigoplus_{R_\pd}\left[\left(\bigotimes_{l \in L_\pd}r_l\right)
\otimes H^{\prime}_A(R_\pd)\right] \,,
\end{equation}
where $H^{\prime}_A(R_{\partial})$ is a bulk Hilbert space spanned by states $|R_A, I_A\rag$. In this decomposition, we use
\begin{equation}
|S_A\rag = |R_\pd\rag\otimes|M\rag\otimes|R_A,I_A\rag \,, \qquad
|S'_A\rag = |R_\pd\rag\otimes|M\rag\otimes|R'_A,I'_A\rag \,.
\end{equation}
Therefore, their outer product is
\bea
|S_A\rag \lag S^{\prime}_A|=|R_{\partial}\rag\lag R_{\partial}|\otimes |M\rag\lag M|\otimes| R_A, I_A\rag\lag R^{\prime}_A, I^{\prime}_A|.
\eea
Using the outer product and rearranging, the reduced density matrix then becomes
\begin{align}
\rho_A 
= \sum_{R_\pd}p(R_\pd)|R_\pd\rag\lag R_\pd|\otimes
\left(\sum_M\frac{|M\rag\lag M|}{\underset{l \in L_{\pd A}}\prod\mathrm{dim}(r_l)}\right)\otimes
\rho_A(R_\pd) \ .
\end{align}
This reduced density can also equivalently be rewritten as
\begin{align}
\bigoplus_{R_\pd}p(R_\pd)
\left[\left(\bigotimes_{l \in L_\pd}\frac{1_{r_l}}{\mathrm{dim}(r_l)}\right)\right]
\otimes\rho_A(R_\pd) \,,
\label{eq:rdmrep}
\end{align}
where 
\begin{equation}
p(R_\pd) = \sum_{R_A,R_{\bar{A}},I_A,I_{\bar{A}}}|\psi(S)|^2 
\end{equation}
is a probability distribution of $R_\pd$ and 
\begin{equation}
\rho_A(R_\pd) = \sum_{R_A, R^{\prime}_A,R_{\bar{A}}, I_A, I^{\prime}_A, I_{\bar{A}}}\frac{\psi(S)\psi(S^{\prime})^*}{p(R_\pd)}
|R_A,I_A\rag\lag R'_A,I'_A|
\end{equation}
is the reduced density matrix. The factor $p(R_\pd)$ is included in the definition to maintain the unit trace condition.

With $\rho_A$ given in \eqref{eq:rdmrep}, the computation of the entanglement entropy can be simplified by using the following properties of the Von-Neumann entropy:
\begin{enumerate}
 \item $S\left(\bigoplus_n p_n\rho_n\right) = H_{C}(p_n) + \lag S(\rho_n)\rag$,
 \item $S(\rho_1\otimes\rho_2) = S(\rho_1) + S(\rho_2)$,
 \item $S(\mathbb(1)_n/n) = \log n$.
\end{enumerate}
Applying these properties to $\rho_A$~\eqref{eq:rdmrep} then yields
\begin{equation}\label{eq:EEELC}
S(\rho_A) = H_{C}(p(R_\pd)) + \sum_{l \in L_{\pd A}}\lag\ln\mathrm{dim}(r_l)\rag + \lag S(\rho_A(R_\pd))\rag \,.
\end{equation}

\end{document}